\documentclass{aa}

\usepackage{graphicx}
\usepackage{txfonts}
\usepackage{soul}
\usepackage{hyperref}
\hypersetup{
    colorlinks   = true,
    urlcolor     = cyan,
    linkcolor    = blue,
    citecolor   = blue
}
\usepackage{threeparttable}
\usepackage{placeins}
\usepackage{xcolor}

\newcommand{\lisca}[1]{LISCA~#1\,}
\newcommand{\mean}[1]{\langle #1 \rangle}

\newcommand{\kms}{km s$^{-1}$\,}
\newcommand{\masyr}{mas yr$^{-1}$\,}
\newcommand{\pmra}{\mu_{\alpha *}}
\newcommand{\pmdec}{\mu_{\delta}}
\newcommand{\vlos}{v_{\rm LOS}}
\newcommand{\rh}{R_{\rm 50}}
\newcommand{\w}{W345\,}
\newcommand{\vRsigmaR}{\langle v_{\rm R}\rangle / \sigma_{\rm R}}
\newcommand{\vx}{V_{\rm X}}
\newcommand{\vy}{V_{\rm Y}}
\newcommand{\vz}{V_{\rm Z}}

\begin{document}

\title{Tracing the W3/W4/W5 and Perseus complex dynamical evolution with star clusters}

\author{
    A. Della Croce \inst{1,2}\thanks{\email{alessandro.dellacroce@inaf.it}}
    \and E. Dalessandro \inst{2} \and 
    E. Vesperini \inst{3} \and
    M. Bellazzini \inst{2} \and
    C. Fanelli \inst{2} \and L. Origlia \inst{2} \and N. Sanna\inst{4}
}

\institute{
    Department of Physics and Astronomy ‘Augusto Righi’, University of Bologna, via Gobetti 93/2, I-40129 Bologna, Italy
    \and
    INAF – Astrophysics and Space Science Observatory of Bologna, via Gobetti 93/3, I-40129 Bologna, Italy
    \and 
    Department of Astronomy, Indiana University, Swain West, 727 E. 3rd Street, IN 47405 Bloomington, USA
    \and 
    INAF - Osservatorio Astrofisico di Arcetri, Largo Enrico Fermi 5, I-50125 Florence, Italy
}

\date{Received \dots; accepted \dots}
 
\abstract
{The Perseus complex offers an ideal testbed to study cluster formation and early evolution as it hosts two major hierarchical structures (namely \lisca{I} and \lisca{II}) and the W3/W4/W5 (W345) region characterized by recent star formation.
This work aims to provide a full characterization of the population of star clusters in the  \w region, in terms of their structural, photometric, and kinematic properties. 
Clusters are then used to probe the dynamical properties of the W345 region and, on a larger scale, to investigate the evolution of the Perseus complex.
We used {\it Gaia} DR3 data to search for star clusters in the \w region and characterize them in terms of their density structure, ellipticity, internal dynamical state, and ages. We also used young stellar object (YSO) catalogs from near-infrared surveys cross-matched with Gaia data to probe their kinematics in the region. 
We identified five stellar clusters belonging to the \w complex. The three younger clusters are still partially embedded in the gas and show evidence of expansion, while the older ones cleared the surrounding gas. We also found that YSOs trace the parent gas structure and possibly its kinematics. 
Thanks to the 6D information available for star clusters, we followed their orbital evolution to assess the formation conditions and evolution of the complex.
When accounting for the Galactic potential, we find that the Perseus complex is not dispersing.
The observed expansion might be a projection effect due to stars orbiting the Galaxy at different velocities.
In addition, we find that the \lisca{I} and \w systems formed some $20-30$~Myr ago just a few hundred parsecs away, while \lisca{II} was originally $\simeq 0.75-1$~kpc apart.
Finally, we also assessed the impact of spiral arm perturbations by constructing tailored Galactic potential which matches the observed Galactic spiral arm structure. We find spiral structures drag star clusters toward higher-density regions, possibly keeping clusters closer for longer than the unperturbed, axisymmetric case.
}

\maketitle

\section{Introduction\label{sec:intro}}
Clustered star formation has been an important mode of star formation since the early Universe. In fact, it is widely accepted that most stars in galaxies
(up to $90\%$ in our Galaxy) form in groups, and spend some time gravitationally bound with their siblings
while still embedded in their progenitor molecular cloud \citep[e.g.,][]{lada_lada2003}.
The vast majority of such systems will be disrupted in their first few million years of existence, due to mechanisms possibly involving gas loss driven by stellar feedback \citep{kroupa_etal2001,baumgardt_kroupa2007,moeckel_bate2010,pelupessy_portegiesZwart2012,pfalzner_kaczmarek2013,brinkmann_etal2017} or encounters with giant molecular clouds \citep[GMCs,][]{gieles_etal2006}. 
Nonetheless, a fraction of these systems will survive the embedded phase and remain bound over longer timescales.

The process of clustered star formation has thus major implications in many fundamental astrophysical
areas, including 
{\it i)} the role of stellar feedback and gas expulsion on the cluster disruption; 
{\it ii)} the contribution of cluster formation to large-scale structures in their host galaxies; 
{\it iii)} the onset of the cluster emerging properties and their evolution; 
{\it iv)} the formation and retention of gravitational wave sources (\citealt{diCarlo_etal2019}), up to {\it v)} the formation of multiple populations \citep[e.g.,][]{bastian_lardo2018,gratton_etal2019}.

Star clusters are expected to form within GMCs with possibly significantly different properties, likely linked to the early organization of the clouds and environmental factors \citep{lada_lada2003,longmore_etal2014}.
In particular, large GMC complexes are expected to produce a large number of star clusters born with a variety of energy distributions \citep{blaauw_1964} and potentially organized in hierarchical multi-scaled structures that can scale down to the formation of compact and bound clusters \citep{dalessandro_etal2021_lisca1,dellacroce_etal2023_lisca2}.
In general, the formation of young stellar clumps and clusters, along with the expansion and dispersal of young star systems may be related to complex dynamical interactions between clusters and sub-clusters during their formation process, and also to the rapid removal of gas due to massive star winds and more violent events like Wolf–Rayet outbursts or supernova explosions that can have an impact on scales of several tens of pc.
Precise astrometric information and accurate line-of-sight (LOS) velocities from the {\it Gaia} satellite and large ground-based spectroscopic surveys have allowed the exploration of star kinematics in recently formed stellar complexes.
In particular, several studies provided evidence of coherent motion and expansion in young star clusters, OB associations, and larger stellar complexes \citep[e.g.,][]{wright_mamajek2018,kounkel_etal2018,damiani_etal2019,kuhn_etal2019,kuhn_etal2020,cantat-gaudin_etal2019,lim_etal2020,lim_etal2022,swiggum_etal2021,das_etal2023,dellacroce_etal2024_expansion,wright_etal2024,jadhav_etal2024,sanchez-sanjuan_etal2024}.

In this context, here we focus on 
the W3/W4/W5 region (hereafter W345) of the Perseus complex.
W345 is a well-studied region of recent star formation, containing two giant
H II regions (W4 and W5), a massive molecular ridge with active star formation (W3), and several embedded star clusters \citep{carpenter_etal2000,koenig_etal2008,roman_zuniga_etal2015,jose_etal2016,sung_etal2017}. 
In addition, we previously identified two major cluster aggregates within the Perseus complex, namely \lisca{I} \citep{dalessandro_etal2021_lisca1} and \lisca{II} \citep{dellacroce_etal2023_lisca2}. These systems encompass several star clusters embedded in a more diffuse stellar halo sharing similar properties in terms of 3D position, 3D velocity, ages, and chemical abundances. Their structural and kinematical properties suggest they might form a more massive cluster by merging smaller systems.
The LISCA systems and the \w region lie within the Perseus: a large and massive complex located toward the Galactic anticenter (at about $2-3$ kpc from the Sun), characterized by recent star formation and showing low dispersion in chemical element abundances \citep{fanelli_etal2022b,fanelli_etal2022}. Also, the complex spatially coincides with a spiral arm structure \citep[see e.g.,][]{reid_etal2019,drimmel_etal2024}.

This work aims to carefully characterize the star cluster population of the W345 region, and to use star cluster within the Perseus complex to trace its formation conditions and evolution.
We start by defining the cluster population in the \w region in Sect.~\ref{sec:preliminary_analysis} and 
Sect.~\ref{sec:star_clusters_w345} presents their structural, photometric, and kinematic properties.
In Sect.~\ref{sec:w345complex_properties} we progressively zoom out studying the young star population in the \w region and their connection to star clusters. Finally, Sect.~\ref{sec:perseus_complex} investigates the star cluster population and evolution of the Perseus complex in a broader Galactic framework. We draw our conclusions in Sect.~\ref{sec:conclusions}.

\section{Identifing star clusters in the \w region\label{sec:preliminary_analysis}}
\begin{figure}[!th]
    \centering
    \includegraphics[width=0.49\textwidth]{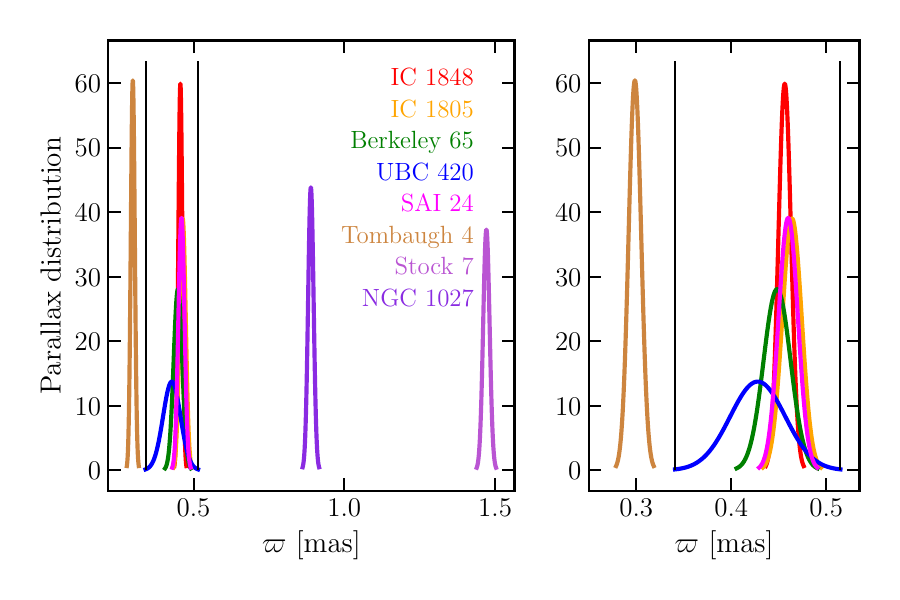}
    \includegraphics[width=0.49\textwidth]{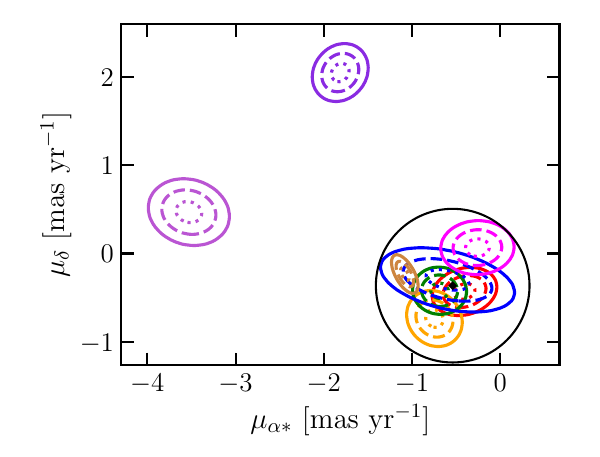}
    \caption{Intrinsic, i.e. deconvolved, parallax (top panel), and PM (bottom panel) distributions for the eight clusters in the region defined by the preliminary Galactic coordinates ranges. Cluster names are reported in the top left panel. Black lines show the parallax and PM ranges adopted for selecting {\it Gaia} sources. 
    The top right panel shows a narrower parallax range centered around the \w star clusters to visualize the cluster parallax distributions better.
    Finally, in the bottom panel, different contours represent the $1\sigma$ (dotted lines), $2\sigma$ (dashed lines), and $3\sigma$ (solid lines) regions for each cluster. Also, correlations among $\pmra$ and $\pmdec$ are visible.}
    \label{fig:par_pm_distribution}
\end{figure}
\begin{figure}[!th]
    \centering
    \includegraphics[width=.5\textwidth]{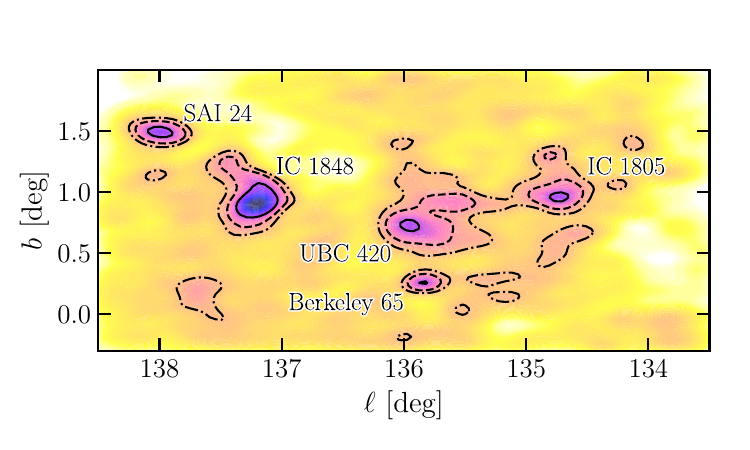}
    \caption{Two-dimensional density map in Galactic coordinates of {\it Gaia} sources after the parallax and PM selections. Darker colors for denser regions. 
    Iso-density contours at $0.5\sigma$ (solid), $1\sigma$ (dashed), and $1.5\sigma$ (dash-dotted) are shown in black.
    The cluster positions and names are marked. 
    The density map was obtained through Gaussian kernel density estimate using the \texttt{gaussian\_kde} function of the \texttt{scipy} Python package \citep{scipy2020}.}
    \label{fig:density_map}
\end{figure}

To study the stellar population of the \w complex we preliminary retrieved from the {\it Gaia} archive\footnote{\url{https://gea.esac.esa.int/archive/}} sources with Galactic coordinates $\ell\in[133.5^{\circ}; 138.5^{\circ}]$, and $b\in[-0.3^\circ; 2^\circ]$.
The ranges are defined such that they enclose the \w complex as traced by the gas distribution at NIR wavelengths \citep[e.g., from the allWISE survey,][]{wright_etal2010}.
We further selected sources with $G$ magnitude brighter than 18 (i.e., $\texttt{phot\_g\_mean\_mag}<18$) and having parallax and proper motion (PM) measurements (i.e., $\texttt{astrometric\_params\_solved}=31$).
We did not apply any prior cut in parallax or PM to avoid possible selection biases in tracing the stellar content (and thus cluster population) of the complex. Instead, we performed data-driven selections as explained in the following \citep[see also][]{dellacroce_etal2023_lisca2}.

We performed an unsupervised clustering analysis on the whole catalog using the $\texttt{HDBSCAN}$\footnote{\url{https://hdbscan.readthedocs.io/en/latest/index.html}} algorithm \citep{McInnes_etal2017_hdbscan} to identify overdensities in the five-dimensional space of Galactic coordinates ($\ell,b$), PM ($\pmra, \pmdec$), and parallax ($\varpi$).
After several tests, we adopted as input parameters $\texttt{min\_cluster\_size} = 50$ and $\texttt{min\_samples} = 30$. 
This combination of parameters allowed us to recover all the known star clusters \citep[e.g.,][]{cantatGaudin_anders_2020,castroGinard_etal2022,hunt_reffert2021,hunt_reffert2023} in the sky region under investigation.
Twelve overdensities were thus identified as star cluster candidates in the region, but four of them were then flagged as false detections by the post-processing routine.
We refer the interested reader to \citet{hunt_reffert2021} for the details of the post-processing routine. Briefly, the nearest-neighbor distance among candidate members and field stars is used as a proxy for local density contrast. Only structures significantly denser than nearby Galactic field stars are considered as true stellar clusters.
The eight clusters identified in our analysis are all known in literature \citep[e.g.,][]{cantatGaudin_anders_2020,hunt_reffert2023}. Namely, they are IC~1848, IC~1805, Berkeley~65, UBC~420, SAI~24, Tombaugh~4, Stock~7, and NGC~1027.
We note that one additional cluster (with a reported number of members greater than 50) is known in the region: UBC~1242 \citep{castroGinard_etal2022,hunt_reffert2023}. Our analysis recovered the cluster but was later rejected as a false positive from the post-processing routine, probably due to the small density contrast with respect to the surrounding field stars.

We then studied the parallax and PM distributions for the eight clusters identified.
The observed parallax and PM distributions were deconvolved using Gaussian modeling and properly accounting for errors and correlations between measurements, thereby obtaining their intrinsic distributions. 
In particular, for each star $i$ we defined the covariance matrix \citep[e.g.,][]{sivia_skilling2006}
\begin{equation}
    \Sigma_i = \begin{pmatrix}
    \delta\pmra^2 & \rho_{\pmra\pmdec} \delta\pmra \delta\pmdec & \rho_{\pmra\varpi} \delta\pmra \delta\varpi \\
    \rho_{\pmra\pmdec} \delta\pmra \delta\pmdec & \delta\pmdec^2      & \rho_{\pmdec\varpi} \delta\pmdec \delta\varpi \\
    \rho_{\pmra\varpi} \delta\pmra \delta\varpi & \rho_{\pmdec\varpi} \delta\pmdec \delta\varpi & \delta\varpi^2
    \end{pmatrix}_i\,,
    \label{eq:cov_matrix_data}
\end{equation}
where $\delta\pmra$, $\delta\pmdec$, and $\delta\varpi$ are the individual errors, and $\rho_{ab}$ is the correlation coefficient among the $a$ and $b$ quantities.
Using the Extreme Deconvolution package\footnote{\url{https://github.com/jobovy/extreme-deconvolution}} by \citet{bovy_etal2011}, we obtained the best fit mean PM components ($\mean{\pmra}$, $\mean{\pmdec}$) and parallax ($\mean{\varpi}$), along with the corresponding dispersions around the mean values, namely $\sigma_{\pmra}$, $\sigma_{\pmdec}$, $\sigma_\varpi$.
We ran this analysis on likely cluster member stars adopting a membership threshold, as defined by the clustering algorithm, of 90\%) and selecting stars with reliable astrometry \citep{lindegren_etal2021}: $\texttt{ruwe}<1.4$, $\delta\varpi/\varpi<0.2$, and $\texttt{astrometric\_excess\_noise}$ smaller than the 95th percentile (applied only to stars with $\texttt{astrometric\_excess\_noise\_sig}>2$).
We tested different membership thresholds (down to 70\%), finding compatible results, within the errors.

Figure~\ref{fig:par_pm_distribution} shows the parallax and PM intrinsic distributions. Some clusters are grouped in both spaces, including the young clusters typically associated with the \w complex, such as IC~1848, and IC~1805.
Therefore, to identify the stars belonging to \w in the {\it Gaia} catalog, we exploited the intrinsic distributions of those clusters showing similar (within $3\sigma$) parallax and PMs. In particular, five out of eight clusters presented compatible distributions in both spaces: IC~1848, IC~1805, Berkeley~65, UBC~420, and SAI~24.
We then defined the sources belonging to the \w region by selecting stars with parallax and PM within $3\sigma$ from the mean value of any of the five clusters.
For the parallax, this translates into sources with $\varpi\in[0.341; 0.515]~$mas. 
In the PM space, sources within $0.87~$\masyr from $(\mean{\pmra},\mean{\pmdec}) = (-0.539;-0.364)~$\masyr were selected. The PM reference point was obtained as the average of the five mean cluster motions. Such ranges are also depicted in Fig.~\ref{fig:par_pm_distribution}.
These data-driven selections in parallax and PM allowed us to adopt some physically motivated ranges for the region under investigation, avoiding any a priori selection. The final catalog counts 8869 sources.

In Table~\ref{tab:mean_cluster_properties} we list the mean cluster properties. We find a good agreement 
($<2\sigma$) with \citet{hunt_reffert2023} for IC~1848, IC~1805, and Berkeley~65 and an excellent ($<1\sigma$) agreement with \citet{cantatGaudin_anders_2020} for UBC~420, which is not included in the "bona fide" cluster sample by \citet{hunt_reffert2023}.

\begin{table*}[!th]
    \centering
    \caption{Mean properties of clusters in the \w complex} \label{tab:mean_cluster_properties}
    \begin{tabular}{lccccccc}
        \hline
        Cluster & $(\alpha, \delta)$ & $\langle\pmra\rangle\quad\sigma_{\pmra}$ & $\langle\pmdec\rangle\quad\sigma_{\pmdec}$ & $\langle\varpi\rangle\quad\sigma_\varpi$ &  $\rho_{\pmra\pmdec}$ & $\rho_{\pmra\varpi}$ & $\rho_{\pmdec\varpi}$\\
        & [degree] & [\masyr] & [\masyr] & [mas] &&&\\
        \hline \hline
        IC~1848     & (42.851,\,60.438) & -0.40\quad0.12 & -0.43\quad0.09 & 0.456\quad0.007 & 0.19 & 0.43 & 0.12        \\
        IC~1805     & (38.198,\,61.510) & -0.75\quad0.11 & -0.74\quad0.11 & 0.464\quad0.010 & -0.13 & -0.32 & -0.09     \\
        Berkeley~65 & (39.766,\,60.408) & -0.69\quad0.10 & -0.42\quad0.09 & 0.448\quad0.014 & -0.03 & -0.60 & -0.19 \\
        UBC~420     & (40.174,\,60.864) & -0.60\quad0.25 & -0.30\quad0.12 & 0.428\quad0.029 & -0.39 & -0.16 & 0.09      \\
        SAI~24      & (44.732,\,60.578) & -0.26\quad0.14 & 0.07\quad 0.10 & 0.460\quad0.010&0.02&0.42&0.30 \\
        \hline
    \end{tabular}
    \begin{tablenotes}
    \item[]\textbf{Notes.} Cluster name, center celestial coordinates, mean PM along the right ascension and declination coordinates, mean parallax, along with the corresponding intrinsic distribution widths.
    The last three columns list the correlation coefficients.
    \end{tablenotes}
\end{table*}

Finally, 
Fig.~\ref{fig:density_map} shows a 2D density map of the region. Clusters clearly appear as overdensities and are clustered in three main regions corresponding to W4 \citep[e.g.,][]{lim_etal2020}, W5-E \citep[e.g.,][]{karr_martin2003}, and W5-W \citep[e.g.,][]{morgan_etal2004}. The W3 region is indeed deeply embedded in the gas and hardly probed by {\it Gaia} \citep[see e.g.,][]{roman_zuniga_etal2015}.

\section{Properties of star clusters in the W3/W4/W5 region\label{sec:star_clusters_w345}}
\subsection{Structure \label{sec:star_clusters_structure}}
\begin{figure*}[!th]
    \centering
    \includegraphics[width=\linewidth]{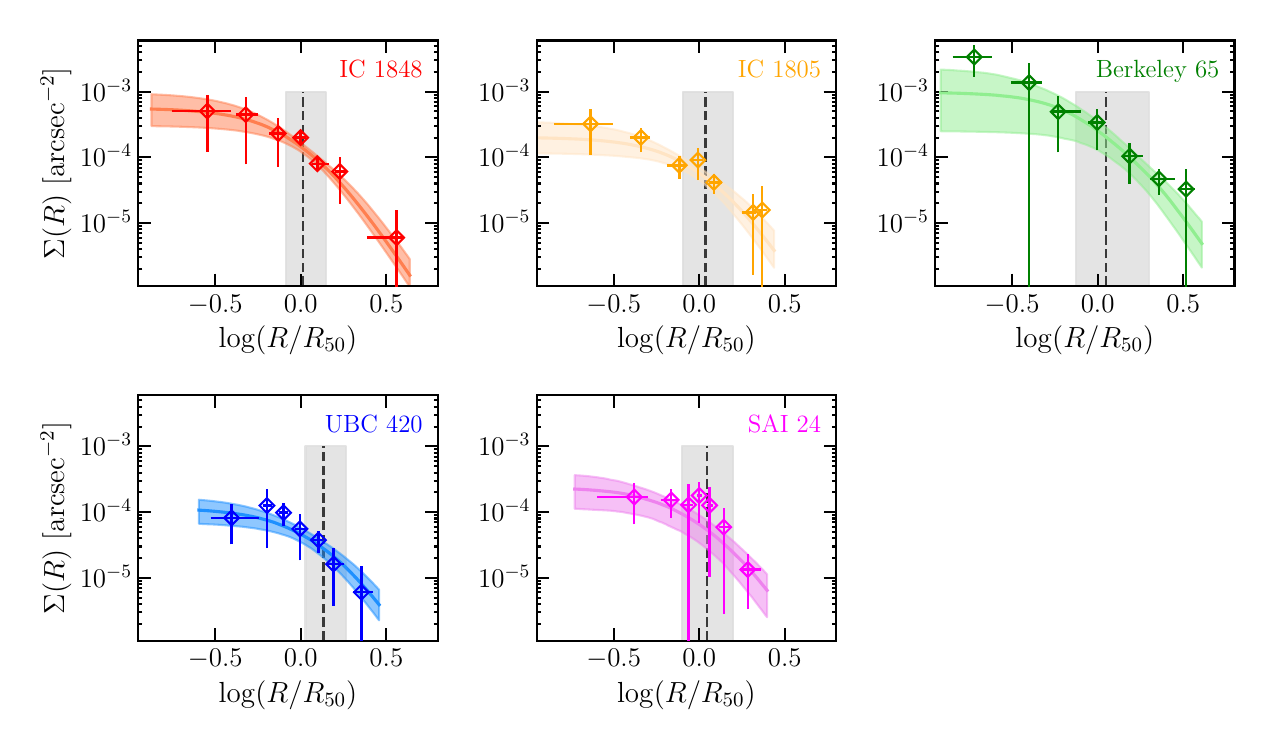}
    \caption{Projected number density profiles for the five clusters analyzed in this study. 
    Distances from the center were normalized to $\rh$.
    Errors in each evenly-populated bin were computed as the standard deviation of density measurements in different angular sectors (concerning the $y$ axis) and as the quantiles of the radial distribution within the bin (for the $x$ axis). The solid lines show the median Plummer model fit of the profile, whereas the shaded areas show the 68\% (i.e., $1\sigma$) confidence interval. Finally, the dashed black lines (the gray shaded areas) show the median Plummer scale radius (68\% confidence interval) from the marginalized posterior distributions.}
    \label{fig:density_profiles}
\end{figure*}
\begin{figure*}[!th]
    \centering
    \includegraphics[width=0.33\textwidth]{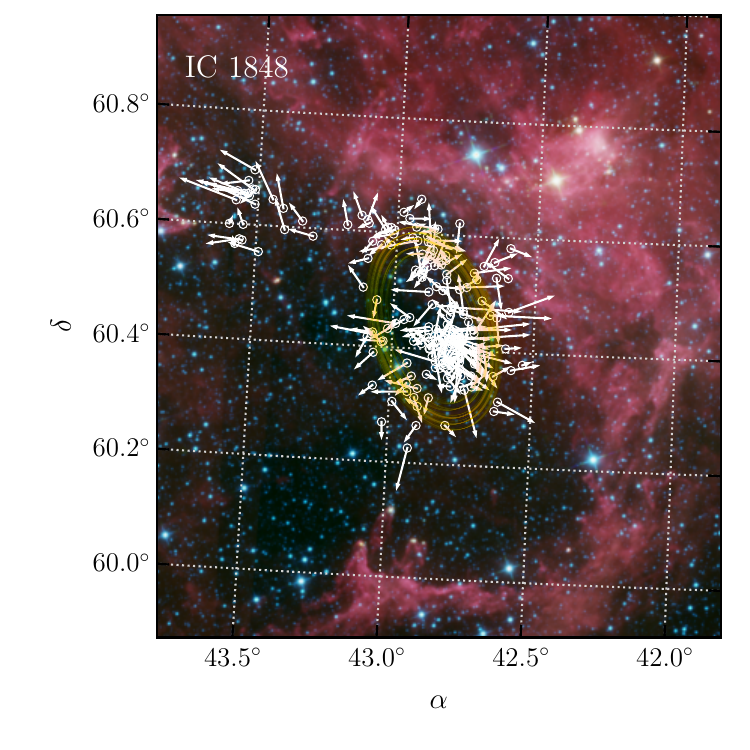}
    \includegraphics[width=0.33\textwidth]{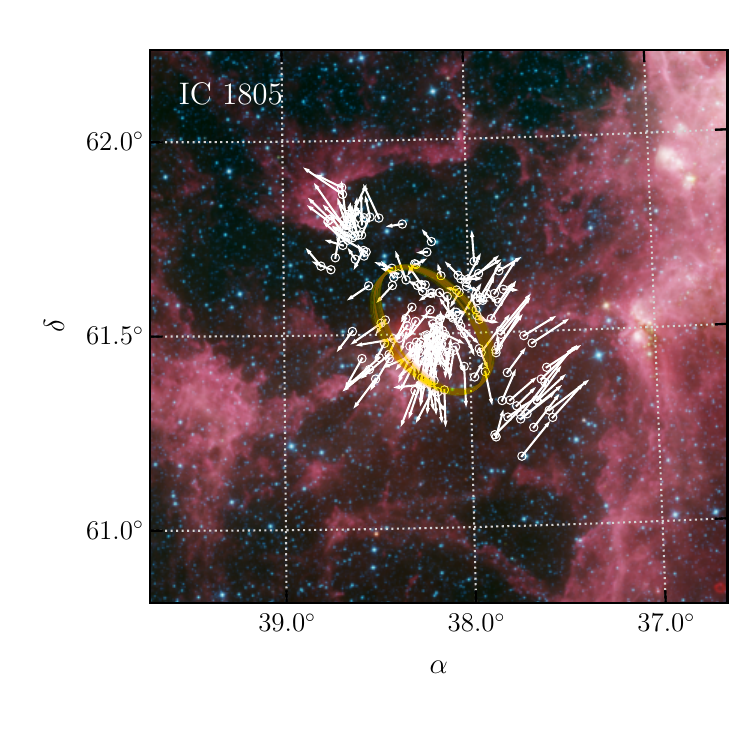}
    \includegraphics[width=0.33\textwidth]{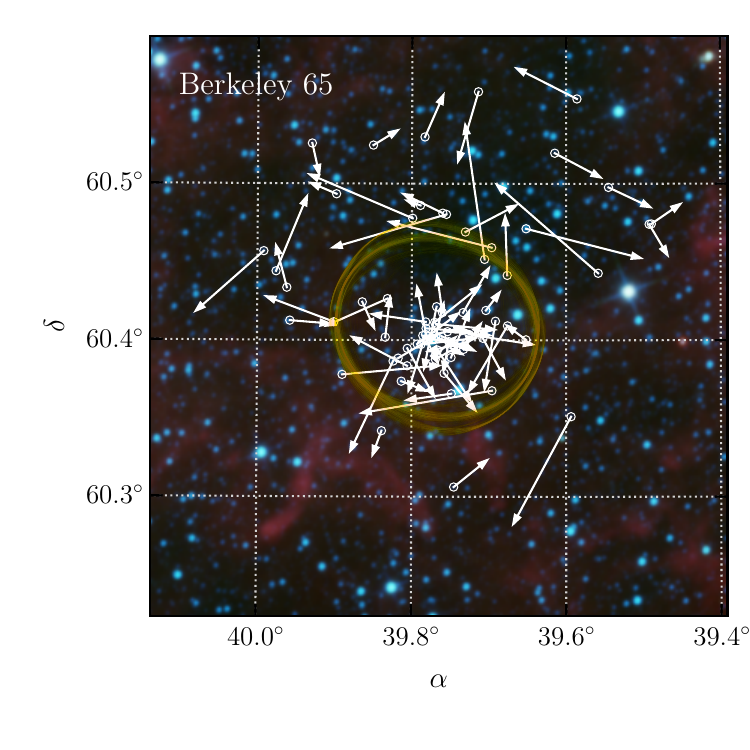}
    \includegraphics[width=0.33\textwidth]{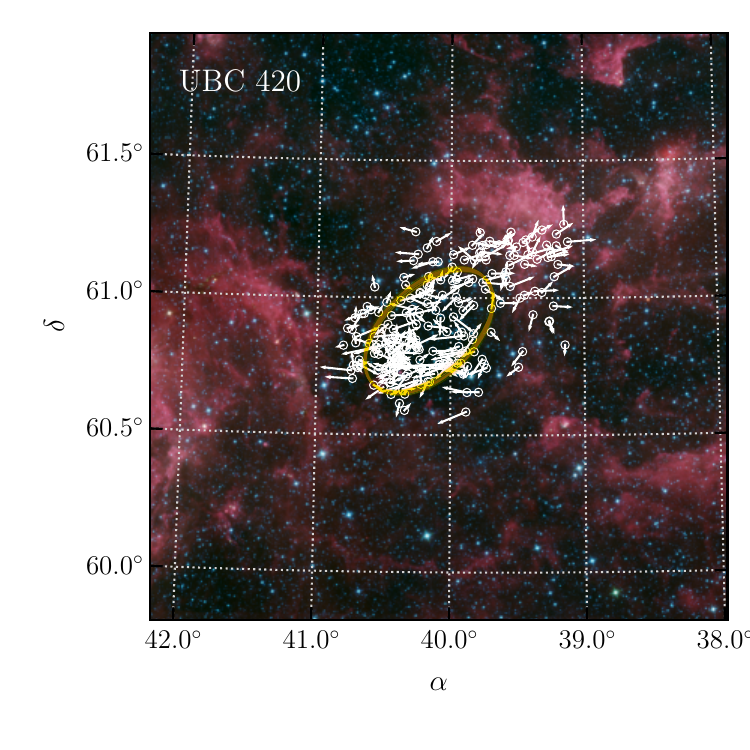}
    \includegraphics[width=0.33\textwidth]{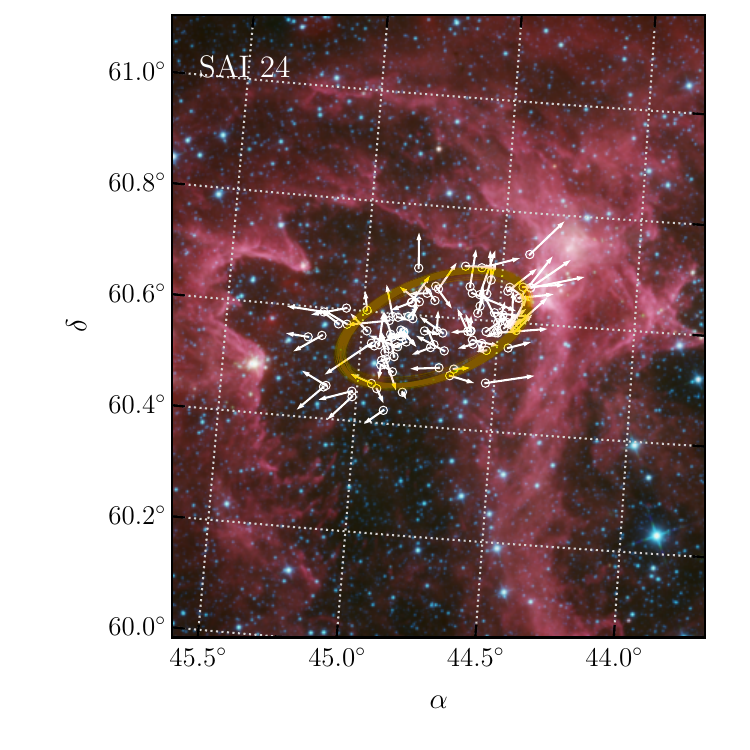}
    \caption{On-sky spatial distribution of cluster members. 
    For each cluster star, PM vectors are shown on top of the RGB image of the region from the allWISE survey.
    We mapped the W3 band in red, W2 in green, and W1 in blue. The W3 filter mainly traces small grain dust and polycyclic aromatic hydrocarbon emissions, whereas the W1 and W2 filters are dominated by young stars \citep{wright_etal2010}.
    The best-fit ellipses of the spatial distributions are shown in gold.} 
    \label{fig:2d_kinematic_maps}
\end{figure*}
\renewcommand{\arraystretch}{1.3}
\begin{table*}[!th]
    \centering
    \caption{Structural, photometric, and kinematic properties of clusters in the \w complex} \label{tab:cluster_structural_properties}
    \begin{tabular}{lcccccc}
        \hline
        Cluster & $R_{\rm Plummer}$ & $\rh$ & $q$ & PA & age & $\vRsigmaR$ \\
        &[arcsec]&[arcsec]& & [deg] & [Myr] & \\
        \hline \hline
        IC~1848     & $377^{+132}_{-77}$    & $367^{ +12 }_{ -10 }$ & $0.59^{+0.02}_{-0.03}$ & $73^{+2}_{-2}$   &$5^{+1}_{-1}$  & $0.64^{+0.08}_{-0.09}$ \\
        IC~1805     & $579^{+264}_{-158}$    & $531^{ +8 }_{ -9 }$ & $0.59^{+0.04}_{-0.07}$ & $51^{+4}_{-4}$   &$5^{+1}_{-1}$  & $1.01^{+0.10}_{-0.08}$ \\
        Berkeley~65 & $151^{+137}_{-51}$  & $133^{ +17 }_{ -2 }$ & $0.85^{+0.05}_{-0.06}$ & $26^{+49}_{-12}$ &$33^{+2}_{-2}$ & $0.06^{+0.13}_{-0.13}$ \\
        UBC~420     & $972^{+355}_{-208}$    & $722^{ +27 }_{ -13 }$ & $0.58^{+0.02}_{-0.04}$ & $137^{+1}_{-2}$  &$8^{+1}_{-1}$  & $0.05^{+0.07}_{-0.07}$ \\
        SAI~24      & $401^{+169}_{-117}$     & $363^{ +5 }_{ -8 }$  & $0.49^{+0.01}_{-0.02}$ & $159^{+2}_{-2}$  &$5^{+1}_{-1}$  & $0.83^{+0.12}_{-0.12}$ \\  
        \hline
    \end{tabular}
    \begin{tablenotes}
    \item[]\textbf{Notes.} Cluster name, Plummer scale radius, median cluster-centric distance, axis ratio, position angle, age, and the ratio between mean velocity and velocity dispersion. Errors on $\rh$, $q$, and PA were computed through 1000 bootstrap extractions of the 90\% of the sample. We stress that such analysis accounts only for fluctuations due to small statistics or outliers. Values obtained through a proper exploration and sampling of the parameter space, as the one performed for $R_{\rm Plummer}$, are thus more realistic. 
    \end{tablenotes}
\end{table*}

We constructed projected number counts density profiles for each cluster using stars with membership above $70\%$
\footnote{For SAI~24 we adopted instead a membership threshold of $75\%$ as a lower threshold would also include small clumps of spatially detached stars, possibly biasing the calculation of the cluster density structure.}.
The density was obtained in spherically symmetric and evenly-populated bins centered on the median cluster positions (see Table~\ref{tab:mean_cluster_properties}).
In each radial bin, the density was computed as the average of the values obtained in 
four angular sectors and the standard deviation of these different measurements (summed in quadrature with Poissonian error) was adopted as the error. 
Figure~\ref{fig:density_profiles} shows the cluster number density profiles obtained this way.
We then fitted the observed density profiles with \citet{plummer_1911} models (see Fig.~\ref{fig:density_profiles}).
The best-fit models were obtained through a Markov chain Monte Carlo (MCMC) exploration of the parameter space, 
using the \texttt{emcee}\footnote{The package is publicly available at \url{https://emcee.readthedocs.io/en/stable/}.} Python package \citep{foreman_mackey_etal2013}.
In general, Plummer models reproduce the observed density profiles fairly well.
Indeed, the Plummer radii ($R_{\rm Plummer}$, inferred from the density profile fitting) and the 2D radii enclosing half of the members ($\rh$, directly computed from the cluster member spatial distributions) agree in general within less than $1\sigma$ (see Fig.~\ref{fig:density_profiles}, and the values reported in Table~\ref{tab:cluster_structural_properties}). 

We further characterized the morphological properties of each cluster by determining the axis ratio ($q$) and the position angle (PA) defined respectively as the ratio between the minor and major axes and the angle between the semi-major axis and the positive $x$ direction.
Following our previous work \citep{dalessandro_etal2024}, we iteratively diagonalized the shape tensor \citep{zemp_etal2011} until a 10\% precision was attained. 
Figure~\ref{fig:2d_kinematic_maps} presents the spatial distribution of cluster members with best-fit ellipses on top of false RGB images of the gas emission in the region. The possible link between elongation and internal kinematics is discussed in Sect.~\ref{sec:star_clusters_kinematics}, while the values are reported in Table~\ref{tab:cluster_structural_properties}.
Generally, all clusters in our analysis are pretty elongated, with the youngest clusters representing the most extreme cases.

\subsection{Differential reddening and cluster ages \label{sec:star_clusters_photometry}}
\begin{figure}[!th]
    \centering
    \includegraphics[width=0.5\textwidth]{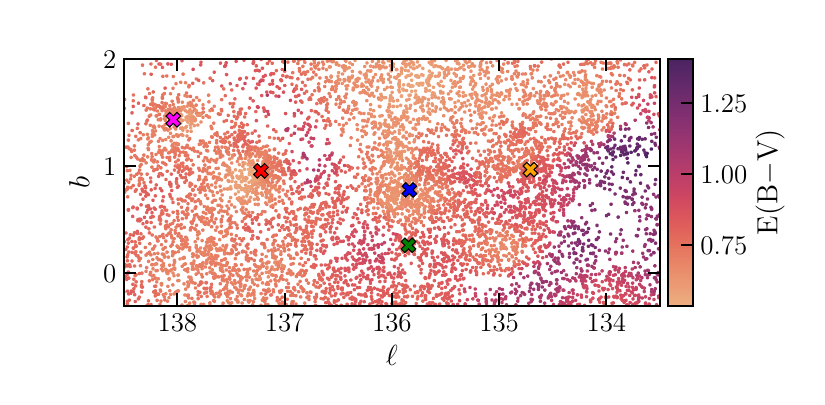}
    \caption{Spatial distribution in Galactic coordinates of {\it Gaia} DR3 sources, selected according to Sect.~\ref{sec:preliminary_analysis}. Each star is color-coded according to its reddening value. Crosses show the centers of the five stellar clusters analyzed in this work (SAI~24 in purple, IC~1848 in red, Berkeley~65 in green, UBC~420 in blue, and IC~1805 in yellow).}
    \label{fig:reddeningmap}
\end{figure}
\begin{figure*}[!th]
    \centering
    \includegraphics[width=\textwidth]{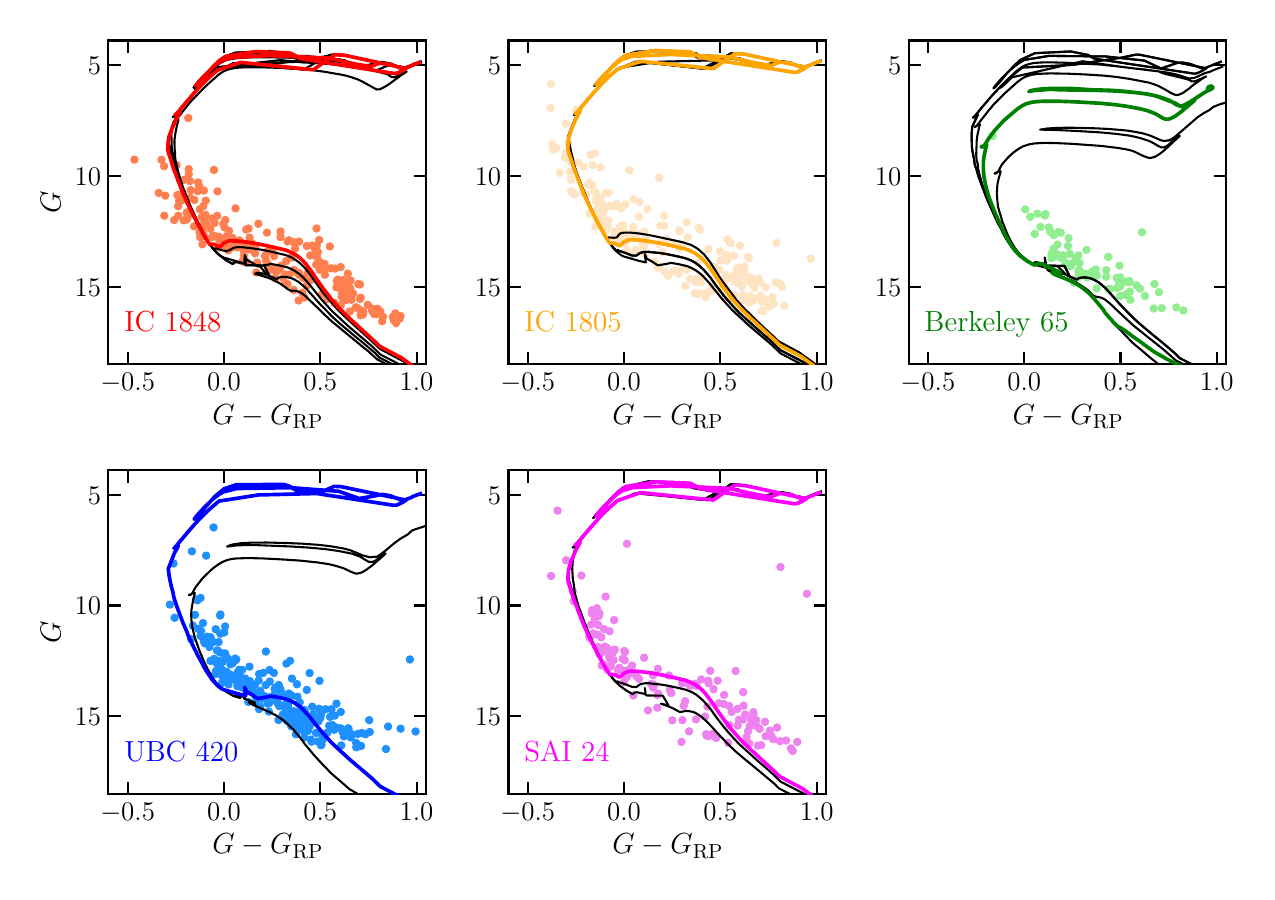}
    \caption{Color magnitude diagrams in the {\it Gaia} filters for cluster members. The best-fit isochrones from differential-reddening-corrected, $G$-band luminosity function fits are shown using the same color palette as cluster members. In black are multiple isochrones from literature works: \citet{catat-gaudin_etal2020_clusterAges,dias_etal2021,hunt_reffert2023,cavallo_etal2024}.}
    \label{fig:isochrone_ageestimates}
\end{figure*}
We constrained cluster ages by fitting the cumulative luminosity function in the differential reddening corrected $G$ band as previously done in \citet{dellacroce_etal2023_lisca2}. 
Differential reddening was computed using the same approach described in the same paper. 
Briefly, {\it Gaia} DR3 sources selected according to Sect.~\ref{sec:preliminary_analysis} were cross-matched with panSTARRS DR1 exploiting the matching tables provided by the {\it Gaia} Data Processing and Analysis Consortium. We thus retrieved $g,r,i,z,$ and $y$-band photometry for the 98\% of the stars selected in {\it Gaia}. We then used the color-color ($G-r$, $i-z$) diagram to assign star-by-star relative reddening values. In particular, for each star, we computed the distance along the reddening vector from the median colors of the closest 50 neighbors (to minimize fluctuations) and a reference point \citep[see][ for further technical details]{dellacroce_etal2023_lisca2}.
We used IC~1848 member stars as a reference, as previous studies \citep[e.g.,][]{cantatGaudin_anders_2020,hunt_reffert2023,cavallo_etal2024} roughly agree on the extinction value of this system, of about $A_{\rm V} = 1.86$ mag. On the contrary, either large discrepancies or much larger extinction values are reported for other clusters, such as Berkeley~65 or IC~1805.
Figure~\ref{fig:reddeningmap} shows the resulting two-dimensional reddening map of the region: as expected, sparser areas are characterized by higher extinction values, suggesting that lower-density regions correspond to areas of significant photometric incompleteness. Indeed, these under-sampled regions trace the spatial distribution of the W3, W4, and W5 GMCs \citep{koenig_etal2008, megeath_etal2008}.

Extinction coefficients also depend on the star effective temperature. This is particularly relevant for wide photometric bands (e.g., the \textit{Gaia} $G$ band) that span a large range of stellar temperatures \citep{jordi_etal2010}.
We accounted for such an effect following \citet{danielski_etal2018} and using extinction temperature-dependent coefficients tailored to {\it Gaia} DR3\footnote{See the online documentation \url{https://www.cosmos.esa.int/web/gaia/edr3-extinction-law}.}.
We note that such relations were calibrated for an effective temperature ($T_{\rm eff}$) range of $3500-10000$~K \citep{Babusiaux_etal2018}, however for clusters as young as 5~Myr we sample stars as hot as $T_{\rm eff} = 30000$~K.
Nonetheless, we adopted such relations for all the stars, although they were extrapolated for the hottest stars in the sample.
Figure~\ref{fig:isochrone_ageestimates} shows the differential reddening corrected color-magnitude diagrams 
along with the best-fit isochrones.
As detailed in \citet{dellacroce_etal2023_lisca2}, cluster ages were constrained using the cumulative luminosity function in the \textit{Gaia} $G$ band. The observed distributions were compared with synthetic simple stellar populations obtained from the PARSEC models \citep{bressan_etal2012_PARSEC}. The method was designed to account for statistical fluctuations in the main-sequence bright-end, possibly due to a combination of physical (e.g., short lifetimes) and instrumental (e.g., saturation) effects. To do so, we randomly picked the observed number of cluster member stars from the synthetic population one hundred times to quantify the probability of missing bright stars due to low numbers.
Our analysis shows that the clusters in the \w region are almost coeval with Berkeley~65 being the older one, with an age of $33^{+2}_{-2}$ Myr. 
Table~\ref{tab:cluster_structural_properties} reports the inferred ages, which are in good agreement with the literature estimates, confirming that IC~1805, IC~1848, and SAI~24 are the youngest clusters in the region (with an age of about 5~Myr). We note however, 
that we found a slightly older age for Berkeley~65 compared to previous estimates (see e.g., \citealt{dias_etal2021,hunt_reffert2023,cavallo_etal2024}, reporting an age in the range $10-17~$Myr, although see also \citealt{catat-gaudin_etal2020_clusterAges}).
Also, for UBC~420 we infer an age of $8^{+1}_{-1}$ Myr significantly younger than the reported age of $\sim 65$ Myr by \citet{catat-gaudin_etal2020_clusterAges}. The reason for the discrepancies may reside in the different member compilations. For example, no UBC~420 members with $G<9~$mag are present in the catalog by \citet{catat-gaudin_etal2020_clusterAges}, possibly leading to the inference of an older age.
Nonetheless, our results confirm that the clusters in the \w region are almost coeval. In this respect, a younger age for Berkeley~65 as suggested by previous studies would further strengthen the conclusion.

\subsection{Kinematics \label{sec:star_clusters_kinematics}}
\begin{figure*}[!th]
    \centering
    \includegraphics[width=\textwidth]{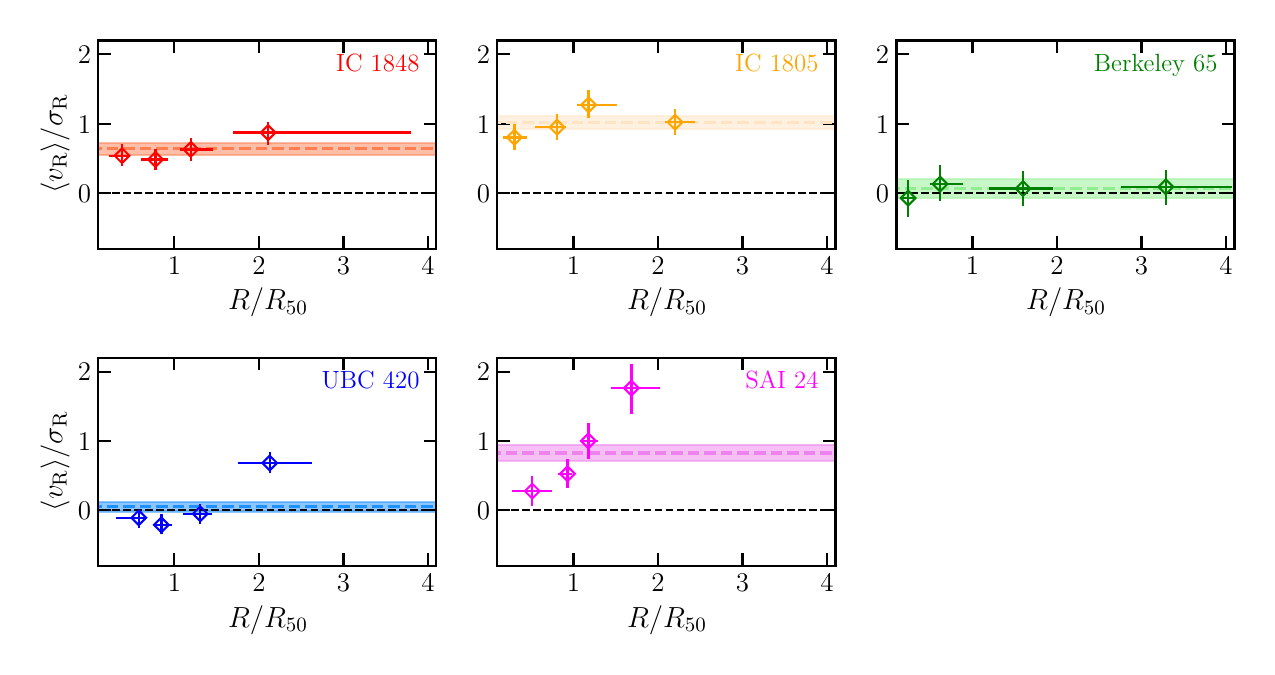}
    \caption{Mean radial velocity to the radial velocity dispersion ratio profiles. Cluster-centric distances were normalized to $R_{\rm 50}$. 
    The integrated values using all the cluster members are shown as horizontal lines (along with errors depicted as shaded areas).
    A black dashed line marks the zero expansion (contraction) level.}
    \label{fig:vRsigmaR_profiles}
\end{figure*}
\begin{figure}[!th]
    \centering
    \includegraphics[width=.5\textwidth]{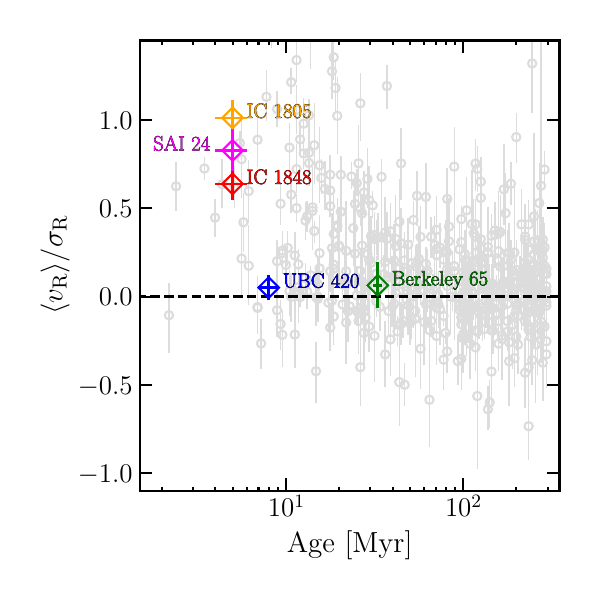}
    \caption{The ratio between the radial mean velocity and velocity dispersion as a function of the cluster age. Underlying data are from \citet{dellacroce_etal2024_expansion}. Different colors highlight the position and results for the clusters analyzed in this work.}
    \label{fig:expansionprop_ageEvolution}
\end{figure}

In this section, we present the analysis of the internal dynamical properties of the five stellar clusters. Their kinematics can in principle give us insights into several physical processes:
from the interplay between gas and stellar dynamics, the role of massive star feedback in both sweeping out the left-over gas and triggering star formation, up to cluster-cluster interactions.

In Fig.~\ref{fig:2d_kinematic_maps} we show the spatial distribution of likely member stars on top of an image of the region using allWISE photometry \citep{wright_etal2010}.
Fig.~\ref{fig:2d_kinematic_maps} shows that some clusters in the sample, namely IC~1848, IC~1805, UBC~420, and SAI~24 are still partially embedded in the gas (at least in projection), whereas Berkeley~65 does not show any clear evidence of surrounding gas. This is consistent with the picture of it being older (see Fig.~\ref{fig:isochrone_ageestimates}), and having completely removed its primordial gas.
Fig.~\ref{fig:2d_kinematic_maps} also shows the projected velocity vectors for member stars: at least three clusters exhibit clear expansion features i.e., IC~1848, IC~1805, and SAI~24, with IC~1805 standing out for amplitude and coherence. Consistently, previous studies reported evidence for expansion for these systems \citep[e.g.,][]{lim_etal2020}.
Several processes can cause clusters to expand, such as left-over gas removal and violent relaxation processes \citep{elmegreen_1983,matiheu_1983,kroupa_etal2001,goodwin_bastian2006, pelupessy_portegiesZwart2012,dinnbier_kroupa2020a,dinnbier_kroupa2020b,dinnbier_etal2022,farias_tan2023,dellacroce_etal2024_expansion}, tidal forces from nearby gas clouds \citep{elmegreen_hunter2010,kruijssen_etal2011}, 
and sub-cluster interactions and mergers \citep[e.g.,][ but see also \citealt{sanchez-sanjuan_etal2024} for an in-depth analysis of the effects of sub-structures on the expansion observed in star-forming complexes]{wright_etal2019}.
In this context, \citet{lim_etal2020} studied the internal kinematics of IC~1805 and found that the cluster is composed of an isotropic core (defined as the region within the half-mass radius) and an external expanding halo.
By investigating the distribution of individual radial velocities\footnote{Throughout this work we refer to "radial velocity" as the PM vector component projected along the radial direction from the cluster center (with the positive direction pointing outward). Where used, we explicitly refer to the velocity obtained from spectra as the line-of-sight velocity.} we found that while a fraction of stars is symmetrically distributed around zero within $0.7~\rh$ (about the half-mass radius reported by \citealt{lim_etal2020}), there is an excess of stars departing from the cluster center with increasing speed, hence driving the expansion signal also within $\rh$ (see e.g. Fig.~\ref{fig:vRsigmaR_profiles}).
Furthermore, most cluster members are enclosed within the tidal radius (roughly corresponding to $2.25~\rh$ according to the estimate by \citealt{lim_etal2020}) and there is no significant evidence of extra-tidal features.
This suggests that internal processes are likely responsible for the observed expansion rather than Galactic tidal forces \citep[as already suggested by][]{lim_etal2020}.

Expansion can also shape the cluster distribution if stars depart faster in one of two orthogonal directions, a process usually referred to as asymmetric expansion \citep[see e.g.,][]{wright_etal2019}. 
In Fig.~\ref{fig:2d_kinematic_maps} we show the ellipses describing the cluster shapes. The PA and $q$ defining these ellipses were obtained as described in Sect.~\ref{sec:star_clusters_structure} (values are reported in Table~\ref{tab:cluster_structural_properties}). All clusters exhibit significant deviation from spherical symmetry and complex structures (as routinely found in many star-forming regions, see e.g. \citealt{cartwright_whitworth2004,gutermuth_etal2008,sanchez-sanjuan_etal2024}).
We thus studied the distribution of individual radial velocities ($v_{\rm R}$, i.e., the proper motion vector projected along the radial direction from the cluster center) as a function of elliptical radii $R_{\rm ell} \equiv \sqrt{ x'^2 + y'^2/q^2}$ (where $x'$ and $y'$ are projected Cartesian coordinates rotated according to PA), or distance along the semi-major axis ($a$). In the case of expansion-driven elongations, we would expect tighter correlations between $v_{\rm R}$ and $R_{\rm ell}$ (or $a$) than with circular radii.
However, we did not find significant differences for all the clusters investigated except possibly for IC~1848.
This suggests that the cluster’s internal kinematics does not drive the present-day cluster morphologies. In contrast, it is likely inherited either from processes that occurred earlier (possibly tidal interactions or mergers) or from the parent gas structure.

Finally, we delved into the expansion features shown by the clusters. In particular, we used the ratio between the mean radial velocity and radial velocity dispersion, $\vRsigmaR$, as defined in \citet[][see Table~\ref{tab:cluster_structural_properties} for the results]{dellacroce_etal2024_expansion}. 
This integrated quantity provides a direct indication of the expansion ($\vRsigmaR>0$), contraction ($\vRsigmaR<0)$, or equilibrium ($\vRsigmaR = 0$) state of the system. Also, it allows meaningful comparison between different clusters as opposed to absolute quantities like $\langle v_{\rm R} \rangle$ \citep[see][]{dellacroce_etal2024_expansion}.
For the five clusters included in this study, we also computed radial profiles of $\vRsigmaR$, presented in Fig.~\ref{fig:vRsigmaR_profiles}, which give us a more complete picture of the cluster's internal kinematics.
Radial trends are observed in IC~1848, IC~1805, and SAI~24, reaching values as high as $\vRsigmaR \simeq 1-2$ (see Fig.~\ref{fig:vRsigmaR_profiles}) while distributions consistent with equilibrium are found in Berkeley~65, and UBC~420.
There are a few more additional interesting points highlighted in Fig.~\ref{fig:vRsigmaR_profiles}: 
Berkeley~65 shows a flat $\vRsigmaR$ profile centered around 0 over a large radial extension ($>3\rh$).
On the other hand, the external stars of UBC~420 are departing from the cluster bulk members. This is particularly evident in the north-east side of the cluster (see Fig.~\ref{fig:2d_kinematic_maps}) possibly suggesting that these stars are currently being stripped from the cluster.

Lastly, we present the expansion properties of clusters in the \w complex within the larger picture of the expansion of young Galactic clusters \citep{dellacroce_etal2024_expansion}. In Figure~\ref{fig:expansionprop_ageEvolution} we show $\vRsigmaR$ as a function of the cluster ages for all the systems studied in \citet{dellacroce_etal2024_expansion} along with the five clusters analyzed in this paper.
The \w complex clusters perfectly fit in the general emerging picture
in which young ($\lesssim 30$~Myr) stellar systems are preferentially expanding, while older ones are roughly compatible with equilibrium configurations.
We also note that except for IC~1848, IC~1805, and SAI~24, other clusters were not included in our previous study due to missing LOS velocity in \citet{tarricq_etal2021} catalog. 
Here we adopted the mean LOS velocity of $-39$~\kms \citep[obtained from high-resolution spectra by][]{fanelli_etal2022}.
Nonetheless, we checked that perspective effect corrections \citep{vanLeeuwen2009} are typically negligible ($<1\%$) for these clusters.

\section{Young stars in the W3/W4/W5 region: their link with star clusters\label{sec:w345complex_properties}}
On a larger scale, the Perseus complex is an extended region of recent star formation located towards the Galactic anti-center, which hosts several star clusters and associations. In particular, some of these clusters were found to be organized in larger hierarchical agglomerates that we named LISCA~I \citep{dalessandro_etal2021_lisca1}, and LISCA~II \citep{dellacroce_etal2023_lisca2}. These are likely the fossils of the star formation within a large gas cloud a few tens of million years ago, and the possible progenitor of massive (a few $10^5 M_{\odot}$) stellar systems forming hierarchically.

\subsection{The YSO population}
\begin{figure*}[!th]
    \centering
    \includegraphics[width=0.75\textwidth]{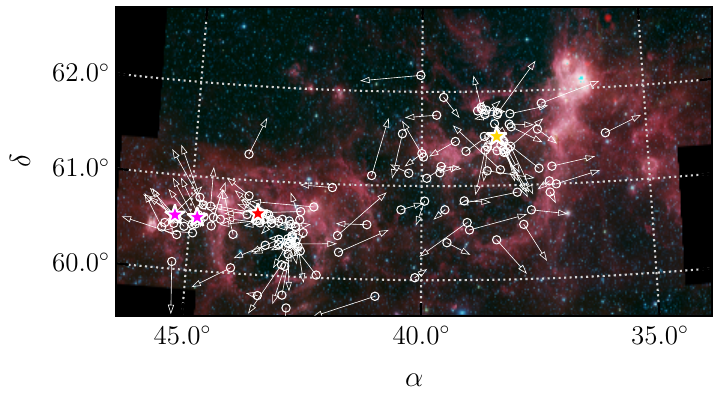}
    \caption{Spatial distribution of the {\it Gaia} YSO sample, with PMs depicted with arrows. 
    PMs were referred to the clusters mean motion in the regions (see Sect.~\ref{sec:preliminary_analysis}).
    The background image is the composite RGB image of the \w complex using data from the allWISE survey. 
    With the star symbols, we show the position of the candidate ionizing sources in the region color-coded according to the star cluster they are assigned to.}
    \label{fig:YSO_kinematics}
\end{figure*}

Young stellar objects (YSOs) trace the most recent star-formation sites, typically younger than a few million years. They thus are used to trace star formation in giant molecular clouds and to constrain the role played by massive stars in either halting or promoting star formation
\citep[see][for a compilation of studies on the \w complex]{massey_etal1995,koenig_etal2008,megeath_etal2008,niwa_etal2009,morgan_etal2009,Chauhan_etal2011,chauhan_etal2011_W5e}.
Many studies looked for YSOs in regions of recent star formation, primarily exploiting color-color selections for their identification \citep[e.g.,][]{allen_etal2004,whitney_etal2004,koenig_etal2008,snider_etal2009,cutri_etal2013_WISE, cutri_etal2021_catalog,yadav_etal2016,jose_etal2016,panwar_etal2017,panwar_etal2019}. 
YSOs indeed exhibit infra-red (IR) emission excess due to infalling, and illuminated material in their early stages and disk emission later on, historically referred to as Class~I and Class~II, respectively, and classified according to their near IR (NIR) spectral energy distribution slope \citep{adams_etal1987,whitney_etal2003b,whitney_etal2003a}.
Also, young stars are prominent X-ray emitters due to magnetic reconnection flares at the stellar surface \citep[see e.g.,][]{gudel_etal2007}, hence X-ray observations could be used to complement infra-red catalogs of YSOs \citep{hofner_etal2002,Feigelson_townsley2008,rauw_naze2016,townsley_etal2019}.

\citet{koenig_etal2008} studied the star-formation history of the \w region, 
by using YSOs identified through the Spitzer space telescope. Furthermore, \citet{panwar_etal2017,panwar_etal2019} characterized the low-mass YSO population around IC~1805 through a multi-wavelength approach from the NIR to the X-rays. In particular, \citet{panwar_etal2019} supplemented the YSO catalog with {\it WISE} data. They used the YSO catalog by \citet{cutri_etal2013_WISE, cutri_etal2021_catalog} from the allWISE program.

While these studies deeply characterized the YSO spatial distribution and its link with the gas, YSO kinematics in the region remains largely unexplored. We thus aim to study their kinematics in the context of the \w complex.
To do so, we cross-matched the catalogs by \citet{koenig_etal2008,cutri_etal2013_WISE, cutri_etal2021_catalog,panwar_etal2019} with {\it Gaia} data.
We selected Class~I and Class~II sources from \citet{koenig_etal2008}, and non-extended, non-variable sources from \citet{cutri_etal2013_WISE, cutri_etal2021_catalog}, with confusion flags "000" and with photometric quality flags not worse than "B" in each band. The starting YSO catalog comprised 4096 sources almost half of which were from \citet{koenig_etal2008}.
The catalog was then crossmatched with {\it Gaia} DR3 sources selected in Sect.~\ref{sec:preliminary_analysis}, obtaining a final catalog of 178 (about 4.4\% of the starting one, out of which 129 had good astrometry according to {\it Gaia} quality flags, see Sect.~\ref{sec:preliminary_analysis}) sources. We will refer to this catalog as the {\it Gaia} YSO.
In Fig.~\ref{fig:YSO_kinematics} we show the spatial distribution of {\it Gaia} YSOs with PM vectors shown as arrows. 
We note that: 
{\it i}) YSOs are spatially concentrated either at the edges of the gas distribution or at young star cluster locations (e.g., SAI~24, IC~1848, IC~1805, see Sect.~\ref{sec:star_clusters_photometry}); 
{\it ii}) YSOs trace the expansion observed in IC~1805, and IC~1848 (see Sect.~\ref{sec:star_clusters_kinematics}); 
{\it iii}) in the W5-E region (around $\alpha \simeq 45^\circ$, see \citealt{karr_martin2003}), YSOs are coherently moving northward. Given their position on the north side of the W5-E HII region, their motion may be inherited from the parent gas.
This analysis further confirms that YSO kinematics is a powerful tool for tracing young cluster formation.

\subsection{Bright-rimmed cloud ionizing sources}
\renewcommand{\arraystretch}{1.3}
\begin{table*}[!th]
    \centering
    \caption{Properties of the ionizing sources in the \w complex} \label{tab:ionizing_sources_BRCs}
    \begin{tabular}{lccccc}
        \hline
        SIMBAD identifier & {\it Gaia} DR3 \texttt{source\_id} & region & BRCs & cluster & membership \\
        \hline \hline
        BD 60 502 / HD 15558 & 465528726379402112 & W4 & 5,7,8,9 & - & - \\
        BD 60 504 / HD 15570 & 465527523789596160 & W4 & 5,6,7,8,9 & - & - \\
        BD 60 507 / HD 15629 & 465535048571192704 & W4 & 5,7,8,9 & IC~1805 & 100\% \\
        HD 17505 & - & W5-West & 10,11 & - & - \\
        BD 60 586 & 464697873547937664 & W5-West & 12 & IC~1848 & 65\% \\
        BD 59 578 / HD 18326 & 466127062559750528 & W5-East & 13,14 &SAI~24 & 100\% \\
        V 1018 Cas & 463122720055223168 & W5-East & pillars & SAI~24 & 72\% \\
        \hline
    \end{tabular}
    \begin{tablenotes}
    \item[]\textbf{Notes.} Star name (multiple identifiers are listed if present). Unique {\it Gaia} DR3 \texttt{source\_id}, crossmatched from the SIMBAD database (for all but HD~17505 which is not in {\it Gaia} although present in the HIPPARCOS catalog). Region of the \w complex where the star belongs. BRCs (or pillars) the source is likely ionizing and triggering star formation (from \citealt{morgan_etal2004}). The cluster and the corresponding membership probability each source was assigned to, if any.
    We used the nomenclature introduced by \citet{karr_martin2003} for the W5 region.
    \end{tablenotes}
\end{table*}

HII regions expanding in the surrounding gas might trigger star formation, forming the so-called bright-rimmed clouds \citep[BRCs,][]{bertoldi1989,bertoldi_mcKee1990,lefloch_lazareff1994,lefloch_lazareff1995}. The bubble expansion drives shock in the surrounding medium 
possibly resulting in gravitationally unstable and triggering star formation \citep{thompson_etal2004}.
The study of BRCs and whether they are in fact star-forming or not \citep{sugitani_etal1991,sugitani_etal1994} can thus provide valuable insights into the role of massive star feedback. 
In addition, identifying the feedback source is crucial to understanding how star formation proceeded.

In this section, we aim to tag candidate ionizing sources to known star clusters and investigate their kinematics within the cluster.
Firstly, we further confirm using {\it Gaia} DR3 parallaxes that the three regions (W3, W4, and W5) lie at the same distance from the Sun, hence their vicinity is not a projection effect, as already suggested by \citet{xu_etal2006,Hachisuka_etal2006,megeath_etal2008}.

We collected candidate ionizing sources from \citet{morgan_etal2004, Deharveng_etal2012}, and used the SIMBAD database to retrieve their {\it Gaia} DR3 \texttt{source\_id} if present (see Table~\ref{tab:ionizing_sources_BRCs}).
This allowed us to look for each source in our cluster member catalogs. We found that at least one source for each region belonged to the central cluster.
BD~60~507 is one of the candidate ionizing sources in W4, possibly triggering star formation in the BRCs 5, 7, 8, and 9 \citep{morgan_etal2004}. It belongs to the star cluster IC~1805 with a high membership probability (see Table~\ref{tab:ionizing_sources_BRCs}). From {\it Gaia} data, we found that the star is close to the cluster center (about $0.5 \rh$) and it is moving at about 0.1~\masyr (i.e., roughly 0.95~\kms at 2~kpc) relative to the mean cluster motion.
Moreover, the star has radial velocity spectrometer spectra, resulting in a fast rotator with $\texttt{vbroad} = 180\pm48~$\kms.

The candidate ionizing source for BRC~12 in the W5-West region is BD~60~586 \citep{morgan_etal2004}. This star was assigned to IC~1848 (with a membership of 65\%). It is located at about $3.6\rh$ but it was not used in the IC~1848 kinematic characterization as it has $\texttt{ruwe} = 2.74$, suggesting that single-star track did not provide a good fit to the observed astrometry.
Nonetheless, the generalized stellar parameterizer from photometry (GSP-Phot) provides $T_{\rm eff} = 21635_{-235}^{+139}~$K.

Concerning the W5-East region, BD~59~578 (also known as HD~18326) is widely referred to as the main ionizing source in the region \citep{Chauhan_etal2011}. According to our catalog, it belongs to SAI~24, being located $\sim0.55\rh$ away from the cluster center, with a relative speed of $0.166$~\masyr (i.e, $1.57$~\kms at 2~kpc).
Besides, \citet{Deharveng_etal2012} suggested V~1018~Cas as an additional ionizing source in W5-East for the observed pillars. The source belongs to SAI~24 although with lower membership, 72\%. It lies at about $2\rh$ with a relative speed of 0.41~\masyr ($\sim 4$~\kms at 2~kpc). According to the GSP-Phot, it has $T_{\rm eff} = 25709_{-246}^{+561}$~K.

Interestingly, none of the cross-matched sources was found to depart at a high relative speed from the cluster regardless of their cluster-centric distance and the expanding nature of the clusters (see Sect.~\ref{sec:star_clusters_kinematics}). This further suggests that those massive stars constitute the core of the clusters. 
Figure~\ref{fig:YSO_kinematics} also shows the position of the candidate ionizing sources tagged to the different star clusters in the region.

\section{The kinematics of the Perseus complex\label{sec:perseus_complex}}
\citet{romanzuniga_etal2019} studied the internal dynamics of the Perseus complex using a sample of young stars and {\it Gaia} DR2 data. The authors found a Hubble-like expansion flow in the region, with an estimated rate of 15~\kms~kpc$^{-1}$. As possible explanations, the authors suggested that the observed expansion could be due to supernova explosions in the region, interactions with the spiral arm, or the result of a large unbound stellar association that is expanding \citep{romanzuniga_etal2019}.
Here we further investigate the dynamics of the complex using its star clusters. There are two main advantages in using star clusters: 
{\it i}) the LOS component of the velocity is more widely accessible compared to individual stars, especially
when dealing with luminous hot ones observed through {\it Gaia} RVS \citep{katz_etal2023}; 
{\it ii}) average position and velocity are more precise and reliable as averaged among several member stars.

\subsection{3D cluster positions and velocities}
\begin{table*}[!th]
    \centering
    \caption{Six-dimensional coordinates of the star clusters within the Perseus complex analyzed in this work} \label{tab:cluster_6D_properties}
    \begin{tabular}{lcccccccc}
        \hline
        Cluster & $\alpha_0$ & $\delta_0$ & $\mean{d}$ & $\mean{\pmra}$ & $\mean{\pmdec}$ & $\vlos$ & $N_{\rm LOS}$ & source \\
        & [$^\circ$] & [$^\circ$] & [pc] & [\masyr] & [\masyr] & [\kms] &&\\
        \hline \hline
Berkeley~65 &39.769&60.403&2013$^{+19}_{-18}$&$-0.688^{+0.014}_{-0.014}$&$-0.423^{+0.013}_{-0.014}$ &   $-60.4 \pm 7.0$ & 1 & (1) \\
IC~1805 &38.198&61.469&1982$^{+10}_{-9}$&$-0.741^{+0.015}_{-0.015}$&$-0.735^{+0.015}_{-0.015}$ &        $-43.8 \pm 3.0$ & 18 & (0) \\
IC~1848 &42.816&60.412&2004$^{+9}_{-9}$&$-0.405^{+0.015}_{-0.014}$&$-0.431^{+0.011}_{-0.011}$ &         $-37.0 \pm 3.3$ & 16 & (0) \\
SAI~24 &44.732&60.577&1989$^{+10}_{-9}$&$-0.257^{+0.018}_{-0.019}$&$0.068^{+0.012}_{-0.012}$ &          $-48.5 \pm 5.4$ & 7 & (1) \\
UBC~420 &40.172&60.862&2106$^{+7}_{-7}$&$-0.599^{+0.017}_{-0.017}$&$-0.298^{+0.008}_{-0.008}$ &         --&--& -- \\
NGC~884 &35.513&57.148&2220$^{+15}_{-12}$&$-0.620^{+0.011}_{-0.010}$&$-1.153^{+0.013}_{-0.014}$ &       $-33.7 \pm 4.0$ & 5 & (0) \\
NGC~869 &34.736&57.130&2217$^{+12}_{-11}$&$-0.651^{+0.010}_{-0.009}$&$-1.161^{+0.011}_{-0.012}$ &       $-67.6 \pm 10.4$ & 6 & (1) \\
NGC~957 &38.348&57.559&2155$^{+19}_{-20}$&$-0.303^{+0.016}_{-0.014}$&$-1.132^{+0.012}_{-0.012}$ &       $-37.5 \pm 5.5$ & 2 & (1) \\
Basel~10 &34.879&58.291&2002$^{+19}_{-19}$&$-0.422^{+0.014}_{-0.015}$&$-0.839^{+0.012}_{-0.013}$ &      --&--&-- \\
NGC~654 &26.004&61.883&2699$^{+20}_{-19}$&$-1.140^{+0.007}_{-0.006}$&$-0.336^{+0.006}_{-0.007}$ &       $-20.6 \pm 8.1$ & 3 & (1) \\
NGC~663 &26.574&61.195&2610$^{+13}_{-11}$&$-1.140^{+0.004}_{-0.004}$&$-0.326^{+0.004}_{-0.004}$ &       $-24.9 \pm 15.8$ & 3 & (1) \\
NGC~659 &26.108&60.677&2843$^{+23}_{-26}$&$-0.831^{+0.004}_{-0.004}$&$-0.301^{+0.005}_{-0.004}$ &       $-48.2 \pm 13.0$ & 2 & (1) \\
NGC~581 &23.339&60.664&2468$^{+17}_{-15}$&$-1.387^{+0.006}_{-0.006}$&$-0.597^{+0.005}_{-0.005}$ &       $-38.1 \pm 4.4$ & 2 & (1) \\
Berkeley~6 &27.801&61.059&2702$^{+31}_{-27}$&$-0.900^{+0.009}_{-0.009}$&$-0.527^{+0.009}_{-0.009}$ &    $-88.9 \pm 5.9$ & 2 & (1) \\
Berkeley~7 &28.545&62.367&2597$^{+27}_{-29}$&$-0.980^{+0.006}_{-0.006}$&$-0.220^{+0.007}_{-0.007}$ &    $-48.4 \pm 5.1$ & 1 & (0) \\
Czernik~6 &30.539&62.839&2653$^{+29}_{-30}$&$-1.184^{+0.008}_{-0.007}$&$-0.186^{+0.011}_{-0.013}$ &     $-56.26 \pm 0.03$ & 1 & (0) \\
Riddle~4 &31.868&60.258&2588$^{+26}_{-29}$&$-0.808^{+0.009}_{-0.010}$&$-0.503^{+0.011}_{-0.012}$ &      $-25.3 \pm 20.3$ & 3 & (1) \\
NGC~637 &25.775&64.041&2524$^{+18}_{-17}$&$-1.259^{+0.006}_{-0.007}$&$-0.027^{+0.006}_{-0.006}$ &       --&--&-- \\
        \hline
    \end{tabular}
    \begin{tablenotes}
    \item[]\textbf{Notes.} Cluster name, center equatorial coordinates, distance from the Sun, PM components, LOS velocity, number of stars used in $\vlos$ calculation, and source catalog for the LOS velocity: (0) for \citet{tarricq_etal2021}, and (1) for \citet{hunt_reffert2023}.
    \end{tablenotes}
\end{table*}

We constructed a catalog of star clusters in the Perseus complex 
by gathering the star clusters of the \w region determined in this paper (see Sect.~\ref{sec:star_clusters_w345}), in LISCA~I \citep{dalessandro_etal2021_lisca1}, and LISCA~II \citep{dellacroce_etal2023_lisca2}.
Cluster member catalogs were either presented in Sect.~\ref{sec:preliminary_analysis} (for \w) or in \citet{dalessandro_etal2021_lisca1} and \citet{dellacroce_etal2023_lisca2}.
We obtained mean sky positions, PM components, and distances for all clusters directly from their members (defined by the membership threshold $>90\%$). Also, we adopted homogeneous astrometric selections (see Sect.~\ref{sec:preliminary_analysis}) and analysis.
Cluster on-sky coordinates $(\alpha_0, \delta_0)$ were obtained by averaging the positions of cluster members.
To derive mean PM and distances for each cluster, we sampled the posterior distribution
\begin{equation}
    \ln p(\mean{\boldsymbol{x}}|\boldsymbol{x}_i) = -\frac{1}{2} \sum^N_{i=1} \left((\mean{\boldsymbol{x}} - \boldsymbol{x}_i)\cdot\Sigma_{\rm conv, i}\cdot(\mean{\boldsymbol{x}} - \boldsymbol{x}_i)^{\rm T} + \ln{|\Sigma_{\rm conv, i}|}\right) .
    \label{eq:distance_estimate}
\end{equation}
In Eq.~\ref{eq:distance_estimate}, $\mean{\boldsymbol{x}} \equiv (\mean{\mu_{\alpha *}}, \mean{\mu_\delta}, 1/\mean{d})$ is the array of mean quantities (with the distance in kpc), and $\boldsymbol{x}_i \equiv (\mu_{\alpha *,i},\mu_{\delta,i},\varpi_i)$ is the array of observables for the $i$-th member star. 
Also, the covariance matrix $\Sigma_{\rm conv, i} \equiv \Sigma_{\rm i} + \Sigma_{\rm model}$ with $\Sigma_{\rm i}$ defined in Eq.~\ref{eq:cov_matrix_data} (to account for the non-negligible correlations in the {\it Gaia} astrometric solution), and 
\begin{equation}
    \Sigma_{\rm model} = \begin{pmatrix}
    \sigma_{\pmra}^2 & \rho_{\rm PM} \sigma_{\pmra} \sigma_{\pmdec}  & 0 \\
    \rho_{\rm PM} \sigma_{\pmra} \sigma_{\pmdec} & \sigma_{\pmdec}^2 & 0 \\
    0 & 0 & 0
    \end{pmatrix} ,
    \label{eq:cov_matrix_data}
\end{equation}
where $\sigma_{\pmra}$, $\sigma_{\pmdec}$, $\rho_{\rm PM}$ are the PM dispersions and correlation coefficients.
Using Eq.~\ref{eq:distance_estimate} we can sample the joint posterior distribution in the mean PM components, velocity dispersions, correlation, and cluster distance. We note however that the distance term in Eq.~\ref{eq:distance_estimate} assumes cluster member stars to lie at the same distance (thereby neglecting the cluster depth) and parallax measurements for nearby stars to be independent (although see \citealt{vasiliev_etal2019}) as discussed by \citet{cantat-gaudin_etal2018}. Nonetheless, here we account for the correlation between PM and parallax as discussed in Sect.~\ref{sec:preliminary_analysis}.
We sample the joint posterior distribution (assuming uniform priors for the parameters) with an MCMC approach using the \texttt{emcee} package by \citet{foreman_mackey_etal2013}. Before sampling the posterior distribution, we accounted for the {\it Gaia} DR3 parallax bias following the prescription by \citet{lindegren_etal2021_parallaxbias}.

Concerning the LOS velocity ($\vlos$) component, we merged multiple catalogs. We adopted the LOS velocity from the catalog with the largest number of member stars with LOS measurement between \citet{tarricq_etal2021} and \citet{hunt_reffert2023}, to have a more robust estimate of the mean LOS velocity.
If a cluster had no measurements in these two catalogs, we searched for LOS velocity measurements in \citet{tsantaki_etal2022_sos}.
The only exception to this general approach is the case of SAI~24. For this cluster, \citet{tarricq_etal2021} measured $\vlos = +52\pm22~$\kms (using 8 stars), whereas \citet{hunt_reffert2023} found $\vlos = -48\pm5~$\kms (using 7 stars). The two measurements are highly discrepant ($3.7\sigma$), possibly due to different membership compilations. However, the measurement from \citet{hunt_reffert2023} is closer to the mean LOS velocity of the \w complex (around $-40~$\kms, \citealt{fanelli_etal2022}) it belongs to. We thus adopted the value by \citet{hunt_reffert2023}. 
We note that SAI~24 is not included in the \citet{tsantaki_etal2022_sos} catalog. 
Furthermore, the clusters UBC~420, Basel~10, and NGC~637 had no LOS velocity measurements in any catalog, whereas Berkeley~6 has a significantly lower $\vlos$ in all the catalogs \citep[has already pointed out in][]{dellacroce_etal2023_lisca2}.

Table~\ref{tab:cluster_6D_properties} presents the six-dimensional phase-space information for the Perseus complex star clusters. The values reported for the distance and mean PM components represent the median value of the marginalized one-dimensional distribution, along with the 16th and 84th percentiles quoted as errors. Our results are generally consistent with those reported by the recent compilation of \citet{hunt_reffert2023}, although we find systematically lower values ($\lesssim 50~$pc) for the cluster distances. We verified that the results presented in the following sections do not change if we adopted \citet{hunt_reffert2023} catalog for the cluster data.

\subsection{The projected kinematics}
\begin{figure*}[!th]
    \centering
    \includegraphics[width=0.56\textwidth]{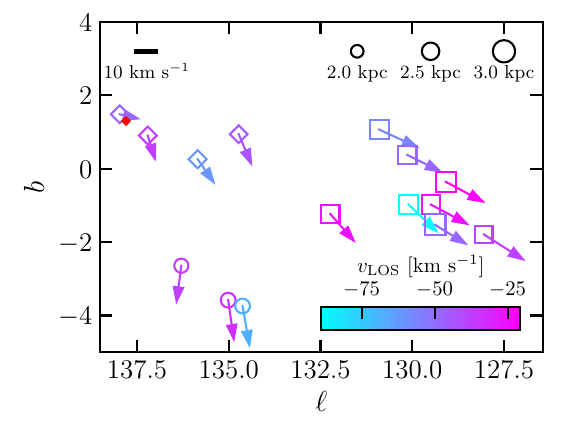}
    \includegraphics[width=0.43\textwidth]{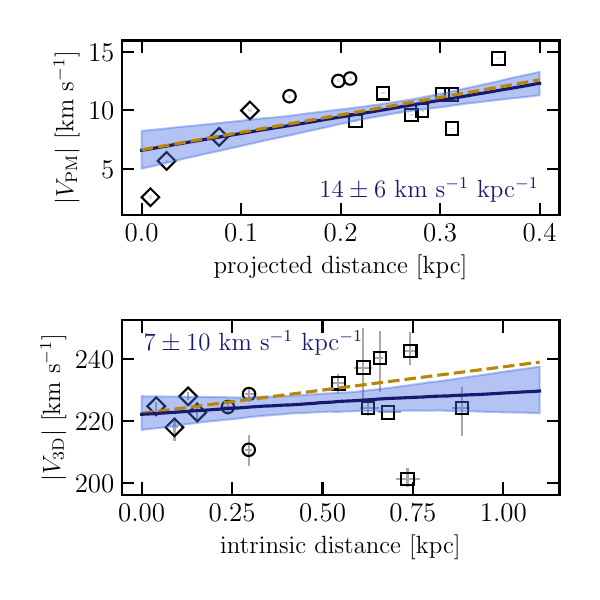}
    \caption{
    Projected kinematic properties for the Perseus star clusters.
    Left panel: spatial distribution of the Perseus complex star clusters. Arrows show the PM vectors, converted in \kms according to the cluster distance (mapped into the size of markers, the smaller, the closest, see the distance coding the top-right corner of the plot). The color coding shows $\vlos$. All data are reported in Table~\ref{tab:cluster_6D_properties}.
    The red diamond shows the claimed expansion origin according to the study of \citet{romanzuniga_etal2019}. 
    Right panels: distance-absolute velocity plots for the projected quantities (top sub-panel) and 3D quantities (i.e., accounting for the position in the Galaxy and $\vlos$ velocity, bottom sub-panel), and 
    assuming that the Sun lies at a distance of $8.178~$kpc \citep{gravitycollab_2019} from the Galactic center, $20.8~$pc above the disk \citep{bennett_bovy2021} and that it is orbiting in the Galaxy at $(\vx,\vy,\vz) = (11.1, 248.5, 7.25)~$\kms \citep{schonrich_etal2010,reid_brunthaler2020}.
    In the top sub-panel, on-sky distances are from the expansion center suggested by \citet{romanzuniga_etal2019}. Velocities and distances were converted in physical units assuming the distance of IC~1805 (see Table~\ref{tab:cluster_6D_properties}). In the sub-bottom panel, 3D distances are relative to IC~1805. Errors are computed by the propagation of distance and $\vlos$ errors (reported in Table~\ref{tab:cluster_6D_properties}).
    Both sub-panels show the linear regression obtained from cluster data (in blue) and the one by \citet[][in brown]{romanzuniga_etal2019}. The best-fit angular coefficient obtained in this work is also reported. Finally, the shaded areas represent the 68\% credible interval, corresponding to $1\sigma$ if the distribution were Gaussian.
    }
    \label{fig:2d_projection_perseusClusters}
\end{figure*}
\begin{figure}[!th]
    \centering
    \includegraphics[width=0.5\textwidth]{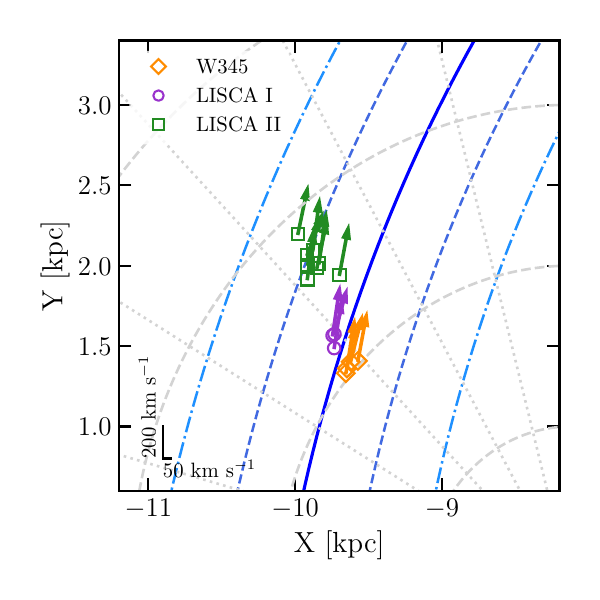}
    \caption{XY projection of the star cluster Galactocentric positions, while arrows show the $\vx$, $\vy$ velocities (with velocity scale reported in the bottom-left corner). Galactocentric coordinates were obtained by direct de-projection of the 6D coordinates listed in Table~\ref{tab:cluster_6D_properties}. Each cluster is color-coded according to the structure it belongs, namely orange diamonds for \w, purple circles for \lisca{I}, and green squares for \lisca{II}. 
    The blue lines show the Perseus spiral arm model (solid) by \citet{reid_etal2019}, with one time (dashed) and twice (dash-dotted) the arm width.
    The background grid is a Heliocentric polar grid with the dashed gray lines showing distances from 1 to 4 kpc, and the dotted ones sampling the angular direction every 15 degrees.
    }
    \label{fig:XYplane}
\end{figure}
In this Section, we present the projected kinematics of the Perseus complex as seen from a star cluster perspective.
Figure~\ref{fig:2d_projection_perseusClusters} shows the absolute, on-sky velocities for the star clusters in the Perseus complex. \lisca{II} clusters are located in $\ell = 128-132$~degree, \lisca{I} lies eastern at $b<-2^\circ$, while in the region $b=0-2$~degrees is \w.
\citet{romanzuniga_etal2019} found a Hubble-like expansion pattern for the region using individual stars. The projected star cluster kinematics qualitatively fits into this picture: looking at the distribution of on-sky absolute PMs as a function of the projected distance we nicely recover the trend reported by \citet{romanzuniga_etal2019}, finding a similar amplitude (see the top-right panel in Fig.~\ref{fig:2d_projection_perseusClusters}).
However, considering the 3D cluster positions and velocities in the Galaxy, this large-scale motion is not observed. By a similar analysis indeed, we found no net trend with the intrinsic distance (see the bottom-right panel of Fig.~\ref{fig:2d_projection_perseusClusters}).
This may be because the Perseus complex spans more than 1~kpc along the LOS, hence when looking at the PM only we are projecting an almost 1.2 kpc-deep on a $10^\circ$-wide region (about 350 pc at 2~kpcs, see left panel of Fig.~\ref{fig:2d_projection_perseusClusters}). 

Given the available 6D data for clusters in the sample (Table~\ref{tab:cluster_6D_properties}), we thus investigated their dynamics in a broader, Galactic framework.
Figure~\ref{fig:XYplane} presents the top-down view of the Galactic disk for the Perseus region. Star cluster positions and in-plane velocities are shown color-coded according to the structure they belong to, along with 
the Perseus spiral-arm model from \citet{reid_etal2019}.
Interestingly, star clusters appear to be moving almost parallel to the arm.
This, together with the Fig.~\ref{fig:2d_projection_perseusClusters} discussion, argues in favor
of the Hubble-like expansion flow reported by \citet{romanzuniga_etal2019} likely being a 
projection effect arising from different orbital velocities at slightly different Galactocentric distances.
In this scenario, the Perseus complex kinematics is governed by the Galactic potential, possibly perturbed by the Perseus spiral arm, rather than internal dynamical processes.

\subsection{Orbits in an axisymmetric potential}
\label{sec:orbits_axysimmetric_potential}
\begin{figure}[!th]
    \centering
    \includegraphics[width=0.24\textwidth]{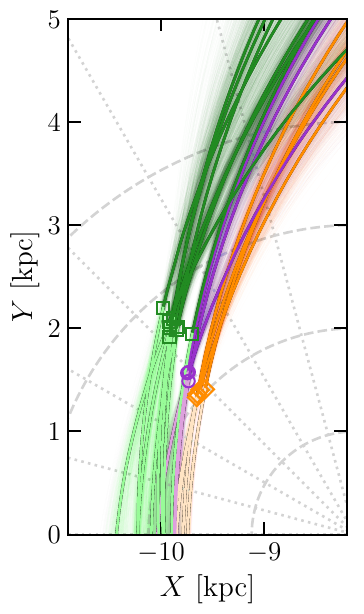}
    \includegraphics[width=0.24\textwidth]{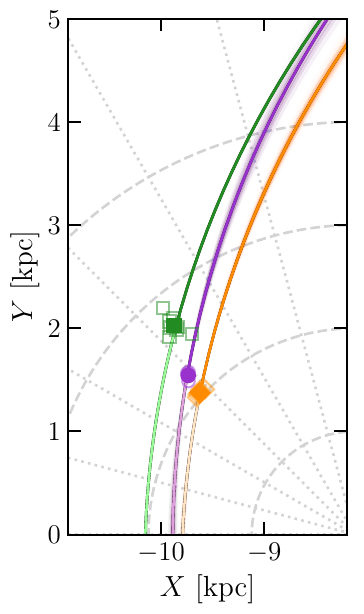}
    \caption{XY projection for individual cluster orbits (left panel) and stellar cluster aggregates (right panel). 
    Darker lines trace the orbits forward in time, whereas lighter ones are backward. 
    Thin lines show orbit integrations from multiple initial condition extractions while thicker ones show median orbits.    
    Present-day cluster positions are also marked: orange diamonds for \w, purple circles for \lisca{I}, and green squares for \lisca{II} clusters.}
    \label{fig:XYprojection_orbits_axisymmetric}
\end{figure}
\begin{figure}[!th]
    \centering
    \includegraphics[width=0.5\textwidth]{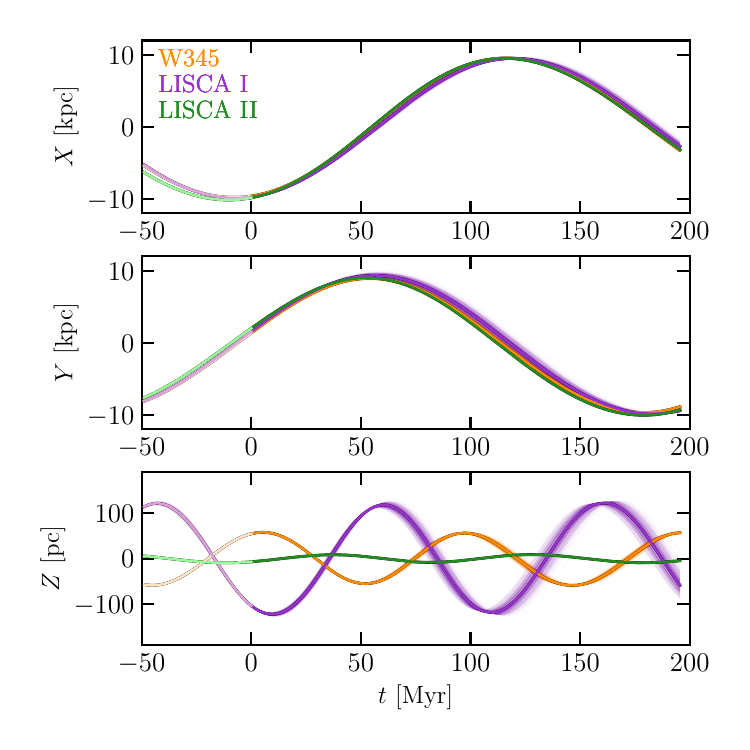}
    \caption{Galactocentric coordinates time evolution for the star cluster aggregates. Darker lines are the forward in time (i.e., $t>0$) integrations while lighter ones are backward (i.e., $t<0$). The present-day positions are at $t=0$. Initial conditions were sampled 500 times to account for errors in the mean distance and $\vlos$, and are shown by thin lines. The thicker solid lines are the median (of those multiple extractions) orbits.}
    \label{fig:xyz_timeEvolution_aggregates_axisymmetric}
\end{figure}
\begin{figure}[!th]
    \centering
    \includegraphics[width=0.5\textwidth]{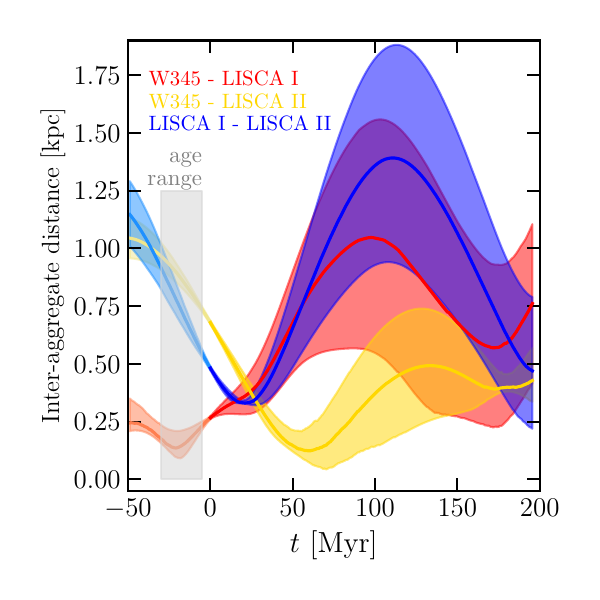}
    \caption{Time evolution of all combinations of inter-aggregate 3D distances. Darker colors trace the forward ($t>0$) integration while lighter colors are the backward integration. Median distances are shown as solid lines, with shaded areas being the 16th (lower distance) and 84th (upper distance) percentiles of the 3D-distance distributions from multiple initial-condition extractions.
    The vertical gray area delimits the stellar age ranges, from 5~Myr (for \w complex star clusters) to about 30~Myr for \lisca{II} \citep[see][]{dellacroce_etal2023_lisca2}.}
    \label{fig:interAggregateDistances_axisymmetricPot}
\end{figure}
We tested the hypothesis that the star clusters in the Perseus complex are not dispersing but rather orbiting in the Galaxy at slightly different Galactocentric distances and speeds, by directly integrating their orbits.
Starting from the 6D projected data (see Table~\ref{tab:cluster_6D_properties}), we obtained Galactocentric coordinates and velocity components assuming that the Sun lies at a distance of $8.178~$kpc \citep{gravitycollab_2019} from the Galactic center, $20.8~$pc above the disk \citep{bennett_bovy2021} and that it is orbiting in the Galaxy at $(\vx,\vy,\vz) = (11.1, 248.5, 7.25)~$\kms \citep{schonrich_etal2010,reid_brunthaler2020}.
Also, since we are measuring cluster mean positions (i.e., mean celestial coordinates and distances), and velocities (i.e., PMs and $\vlos$), literature catalogs provide errors in these quantities \citep[][see Table~\ref{tab:cluster_6D_properties}]{tarricq_etal2021,hunt_reffert2023}.
We thus accounted for distance and $\vlos$ errors (typically the primary uncertainty sources) by sampling their distributions 500 times and obtaining Galactocentric coordinates for all the extractions. 
Orbits were thus integrated in the axisymmetric \citet{mcmillan_etal2017} potential using the Action-based Galaxy 
Modelling Architecture (\texttt{AGAMA}
\footnote{Publicy available at \url{https://github.com/GalacticDynamics-Oxford/Agama}.}) 
library \citep{vasiliev_etal2019_agama} for 200 Myr. 
Besides, to constrain the formation scenario and the initial size of the Perseus complex we performed a backward orbit integration by flipping the velocity vectors for each cluster. 
In the framework of the canonical set of angle-action coordinates \citep{binney_tremaine2008}, this transformation allows us to follow the same orbit (as the action integrals are unchanged) but in the opposite direction along the angle space. Backward orbits were integrated for 50~Myr which is genearlly larger than the stellar ages in the region \citep{dellacroce_etal2023_lisca2,hunt_reffert2023,cavallo_etal2024}.
In Fig.\ref{fig:XYprojection_orbits_axisymmetric} we show the top-down view of the Galactic plane with individual cluster orbits both forward and backward in time. 
Also, in Appendix~\ref{appendix:cluster_orbits} we present and discuss the time evolution of Galactocentric coordinates for every cluster.
There are a few interesting points to highlight from Fig.~\ref{fig:XYprojection_orbits_axisymmetric}: 
{\it i}) \lisca{I} clusters move slightly outward in the Galactic plane, and toward \lisca{II};
{\it ii}) two clusters, namely Berkeley~65 and NGC~869, depart the most from the other neighbor clusters (\lisca{II} and \lisca{I} clusters respectively). This is likely due to the difference in their $\vlos$ to the other clusters (see Table~\ref{tab:cluster_6D_properties});
{\it iii}) going back in time in the cluster orbit reconstruction, \lisca{I} appears to converge toward the \w complex.

Finally, we note that accounting for distance and $\vlos$ errors is key in constraining the origin and evolution of star cluster systems. Although on Galactic scales the orbits are well constrained (see Fig.~\ref{fig:orbits_starCluster_axisymemtricPotential}) when looking at their distribution on cluster scales (or inter-cluster distance scales) errors on the initial conditions are far from negligible. This may also account for the apparent separation of \lisca{II} star clusters into two branches.

While individual star cluster orbits could give us insights into their evolution and origin, they may suffer of strong assumptions and shortcomings, particularly in the context of the Perseus complex. 
Star clusters in \lisca{I} and \lisca{II} are not isolated \citep{dalessandro_etal2021_lisca1,dellacroce_etal2024_expansion}. On the contrary, clusters are likely interacting with each other and they are also embedded in a more diffuse stellar halo.
Therefore their present-day properties (mainly in terms of velocity distributions) may be strongly affected by their mutual interaction with other clusters.
We thus decided to study the Persues complex at large using larger structures, such as the LISCA systems and clusters in the \w region. We shall refer to these structures as cluster aggregates.
Such an approach has multiple advantages. Firstly, it allows us to trace the Perseus complex evolution on finer scales. Secondly, we are not considering star clusters as isolated entities, but rather as part of larger complexes (as suggested by previous studies, e.g., \citealt{dalessandro_etal2021_lisca1}, and \citealt{dellacroce_etal2023_lisca2}).
We stress here, though, that despite the evidence of \w complex star clusters being co-moving and co-spatial, treating them as a single aggregate does not imply any physical connection as for the LISCAs.

We obtained mean Galactocentric positions and velocities for the cluster aggregates by averaging the individual cluster quantities. In particular, we followed \citet{sivia_skilling2006} to estimate mean values and standard errors while taking into account heterogeneous errors.
Finally, Fig.~\ref{fig:XYprojection_orbits_axisymmetric} (right panel) presents the cluster aggregate orbits projected on the XY plane. 
As anticipated, the use of cluster aggregates provides a clearer picture: \lisca{I} is currently in the process of drifting away from the \w complex toward \lisca{II}, while \lisca{II} and \w seems to evolve in parallel.
Furthermore, in Fig.~\ref{fig:xyz_timeEvolution_aggregates_axisymmetric} we present the evolution of $XYZ$ coordinates with time. Interestingly, the three structures evolve similarly in the plane for almost a full orbital period (of about $250-260~$Myr) while experiencing different oscillation amplitude up and down the plane, with \lisca{I} showing the largest amplitude, up to 100~pc.

Lastly, to quantitatively constrain the evolution and formation of the Perseus complex, we traced the 3D inter-aggregate distances with time (Fig.~\ref{fig:interAggregateDistances_axisymmetricPot}).
At each time step, distances between the three cluster aggregates were computed for each orbit from the pool of initial conditions, and the median distance, along with the 16th and 84th percentiles, were obtained.
We conclude that: 
{\it i}) \lisca{I} is currently moving away from \w, reaching a distance of about 1.1~kpc in 100~Myr, before approaching it again; 
{\it ii}) at the same time, \lisca{I} and \lisca{II} are getting closer, reaching minimum distance in about $30-40~$Myr 
{\it iii}) despite their appearance in the $XY$ plane, \lisca{II} and \w are in fact approaching each other. In about 60~Myr they reach a distance of a few hundred parsecs, before slowly departing; 
{\it iv}) concerning the backward integration, we can trace the formation condition of the Perseus complex. \lisca{I} and the \w region were at their minimum distance about 25~Myr ago of just a few hundred parsecs, while \lisca{II} formed further away, between $0.6 - 1~$kpc; 
{\it v}) we do not observe a Hubble-like expansion of the region 
as suggested by \citet{romanzuniga_etal2019}. This highlights the importance of orbit integration (see e.g., Fig.~\ref{fig:interAggregateDistances_axisymmetricPot}) and of using star clusters as tracers of the evolution of the complex.

\subsection{Orbits in a spiral-perturbed potential}
\label{sec:orbits:spiral_potential}
\begin{figure}[!th]
    \centering
    \includegraphics[width=0.5\textwidth]{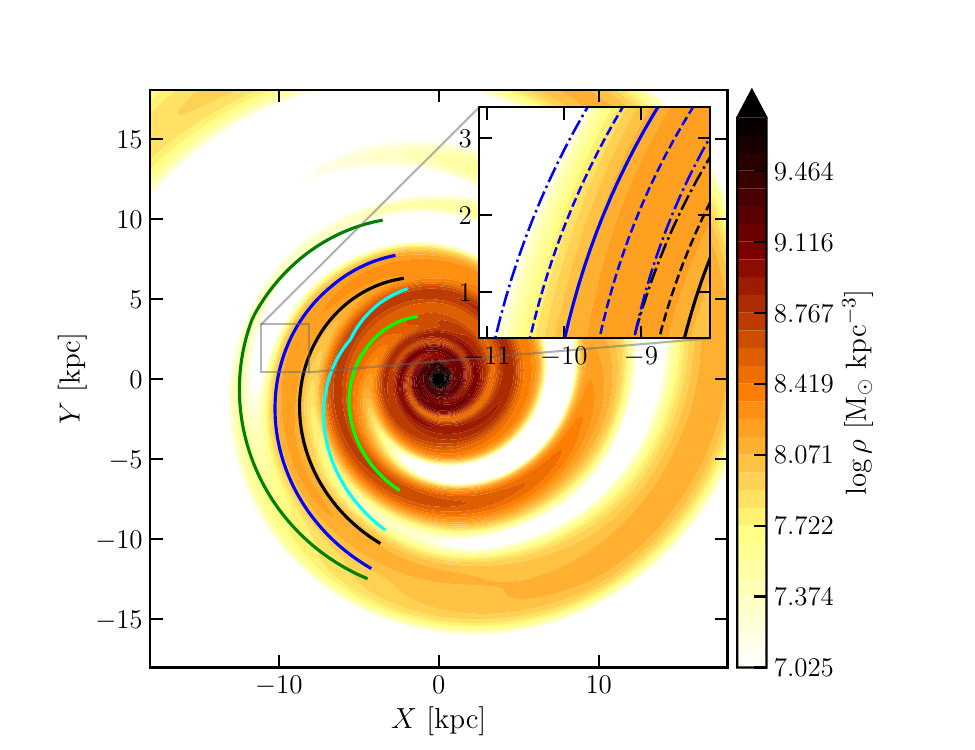}
    \caption{Density map on the XY Galactic plane (i.e., computed at $Z=0$) for a spiral perturbation with $f=0.3$ (used as a reference model). The Sun is located at $X=-8.178~$kpc and $Y=0$. 
    The different colored lines present the different spiral arm models according to \citet{reid_etal2019}: in dark green is the Outer arm, in blue the Perseus one, in black the Local, in cyan the Sagittarius-Carina and in lime the Scutum one. In the inset, we show a zoom-in of the Perseus region.}
    \label{fig:densitymap_f03_spiralpotential}
\end{figure}
\begin{figure}[!th]
    \centering
    \includegraphics[width=0.5\textwidth]{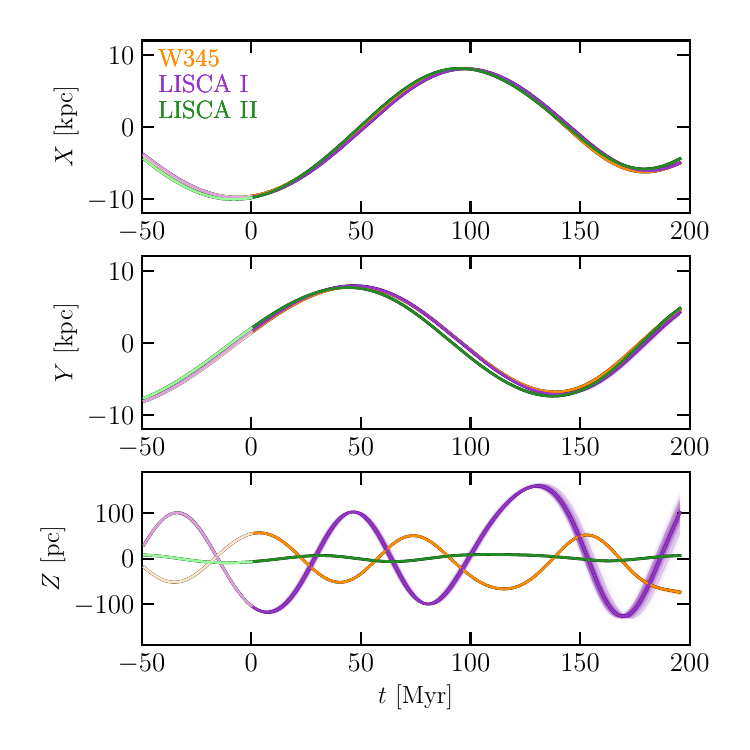}
    \caption{The same as Fig.~\ref{fig:xyz_timeEvolution_aggregates_axisymmetric} but for a spiral-perturbed potential with $f=0.3$.}
    \label{fig:xyzEvolution_f03_spiral}
\end{figure}
\begin{figure}[!th]
    \centering
    \includegraphics[width=0.5\textwidth]{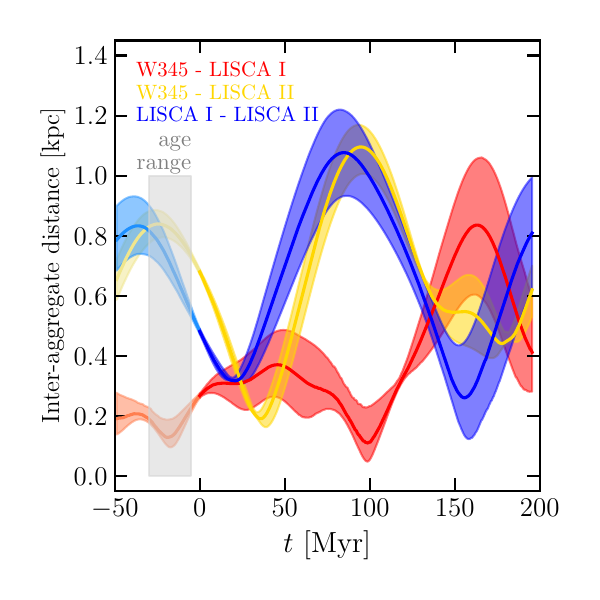}
    \caption{The same as Fig.~\ref{fig:interAggregateDistances_axisymmetricPot} but for a spiral-perturbed potential with $f=0.3$.}
    \label{fig:distanceEvolution_f03_spiral}
\end{figure}

Figure~\ref{fig:XYplane} shows that Perseus star clusters lie on the spiral arm \citep{reid_etal2019}.
Spiral structures \citep[see][for different formation theories]{lin_shu1964,shu_2016,sellwood_carlberg1984} are believed to play an important role in gathering gas, triggering star formation, and perturbing stellar orbits thanks to the locally deeper potential well \citep{baba_etal2016,tchernyshyov_etal2018}. 
In addition, \citet{romanzuniga_etal2019} suggested that the apparent expansion of the Perseus complex could be due to spiral arm interaction.
We thus built a Galactic potential toy model in which we consider spiral arm perturbations to assess their effect on the cluster orbits.

We started from the spiral arm model provided by \citet{reid_etal2019}. The authors modeled the spiral arm shape as
\begin{equation}
    \ln(R/R_{\rm kink}) = -(\theta - \theta_{\rm kink})\tan \psi \,,
    \label{eq:spiralarm}
\end{equation}
where $R_{\rm kink}$, and $\theta_{\rm kink}$ are characteristic Galactocentric radius (on the plane) and azimuth respectively, and $\psi$ is the pitch angle.
Also, they allowed the arm to abruptly change the pitch angle at $\theta_{\rm kink}$ and considered a distance-dependent arm width (see Fig.~\ref{fig:XYplane}).
Finally, \citet{reid_etal2019} concluded that the young-star distribution is consistent with a five-arm model including the Scutum, the Sagittarius-Carina, the Local, the Perseus, and the Outer spiral arm.
We thus used the potential formulation derived by \citet{cox_gomez2002} for the density in Eq.~\ref{eq:spiralarm}, assuming each arm is a single spiral, and considered them as perturbations of the underlying axisymmetric potential \citep{mcmillan_etal2017}.
Such an approach has some assumptions and limitations that we discuss here.
{\it i}) The potential model \citep{cox_gomez2002} does not allow for a change in the pitch angle, we thus assumed the average value for $\psi$. In particular, for the Perseus arm, we assumed $\psi = 9.5^\circ$;
{\it ii}) the amplitude of the density variations ($f$) with respect to the underlying potential is largely unconstrained by observations. This is a rather key parameter as increasing $f$ enhances density perturbations. \citet{levine_etal2006} found that the density ratio between arm and inter-arm region using HI observations is about $\sim 3$. Assuming that the stellar distribution presents the same ratio, we can translate it into $f$, finding that $f_{\rm HI} = 0.53-0.66$. However, this assumption is most likely wrong as the arm-to-inter-arm ratio for the gas is expected to be larger than the stellar one. Indeed, the gas gathers in the overdensities (i.e., the spiral arm) thereby being converted into stars. On the other hand, stars formed in the arms can drift away populating the interarm region. Therefore, $f_{\rm HI}$ represents an upper limit;
{\it iii}) recent studies found that spiral arm pattern speed decreases with Galactocentric radius \citep{naoz_shaviv2007,castro-ginard_etal2021_spiralarm}. Our spiral arm model does not allow for different pattern speeds as orbits are integrated into the non-inertial, co-rotating frame. Since we are mainly interested in the role of the Perseus arm we decided to adopt the pattern speed $\Omega_{\rm p} = 17.82 \pm 2.98$~\kms~kpc$^{-1}$ derived for the Perseus arm by \citet{castro-ginard_etal2021_spiralarm} from the youngest open-cluster sample (that supposedly better traces the spiral structures in which they formed);
{\it iv}) our potential model does not include the bar which was shown to have a prominent impact on the Galactic stellar kinematics \citep[e.g.,][]{kawata_etal2021,drimmel_etal2023}. 
Orbits were integrated for at most 200~Myr, during which clusters orbit at about 10~kpc from the Galactic center. Hence the role of the bar could be treated as a second-order effect compared to the local spiral arm potential perturbations. We nevertheless 
present in Appendix~\ref{appendix:bar_potential} the orbits for a Galactic potential which includes the bar;
{\it v}) spiral arms unrealistically extend to the Galactic center, well beyond the data coverage \citep{reid_etal2019}. However, as discussed above, we are mainly interested in the local effects on the Perseus complex.
Despite the aforementioned caveats, Fig.~\ref{fig:densitymap_f03_spiralpotential} shows the density map projected on the Galactic plane for the perturbed \citet{mcmillan_etal2017} potential with $f=0.3$. Qualitatively, the density potential model closely resembles the spiral arm structure presented by \citet[][see the spiral structures in Fig.~\ref{fig:densitymap_f03_spiralpotential}]{reid_etal2019}.
We explored several $f$ finding qualitatively similar results, and we present them in Appendix~\ref{appendix:different_spiral_models}.

Having built the Galactic potential, we computed the cluster aggregate orbits 
to study the evolution of the Perseus complex in the presence of spiral-arm perturbation. Figure~\ref{fig:xyzEvolution_f03_spiral} presents the evolution of Galactocentric coordinates with time. 
We broadly found that the Perseus spiral arm pulls star clusters towards higher-density regions during their orbit. 
Since the star clusters in the Perseus complex are within the spiral arm, the stronger gravitational force may keep the star clusters closer for large times when compared to the axisymmetric case. Figure~\ref{fig:distanceEvolution_f03_spiral} shows the inter-aggregate distance evolution for the perturbed ($f=0.3$) potential. Opposed to the axisymmetric case (Fig.~\ref{fig:interAggregateDistances_axisymmetricPot}), their relative distances oscillate between about $0.25 - 1~$kpc for more than 200~Myr. 
However, star clusters may escape the spiral arm due to either differences in the orbital frequency and pattern speed or net relative inclination of the velocity vector to the arm pitch angle. In these cases, the Perseus spiral arm (or other nearby arms) drags the star clusters, profoundly changing the orbit. Such orbit perturbations are largely dependent on the value of $f$ (see for instance Fig.~\ref{fig:distanceEvolution_spiralModels_AppendixB}).
Finally, integrating the orbits backward in time\footnote{
To integrate orbits backward also the potential should evolve accordingly. Therefore, the pattern speed sign was changed. We verified that the integrator adopted conserves the Jacobi integral, with a typical relative precision of a few $10^{-8}$ for the investigated temporal range.
} in the presence of spiral arm perturbations produces similar results with the interesting trend of decreasing the distance of \lisca{II} about 30~Myr ago with increasing $f$ down to $500-750$~pc (see Appendix~\ref{appendix:different_spiral_models}). 
This suggests that the Perseus complex could have been smaller in size if the spiral arms were more prominent when its major stellar associations formed.

We conclude that spiral arms play a role in shaping the cluster orbits and in the evolution of the Perseus complex.
Also, different values of $f$ do not qualitatively change the evolution on short time scales $\lesssim 100~$Myr. 

\section{Summary and conclusions\label{sec:conclusions}}
We studied the properties of the W345 region by using its cluster population and in the context of the more extended Perseus Complex.
We identified five clusters in the \w region, namely IC~1805, IC~1848, Berkeley~65, UBC~420, and SAI~24, all previously known in the literature and sharing similar 3D velocities and positions.
All the clusters exhibit well-defined density structures as shown by comparing the observed density profiles with theoretical models. Also, they present significant deviations from spherical symmetry. 
We found no evidence of a link between clusters’ morphological properties and asymmetric expansion thus suggesting the present-day spatial distribution is likely inherited
from earlier processes or star formation.
On the internal kinematics side, the three youngest (IC~1805, IC~1848, and SAI~24) clusters show prominent expansion, consistent with the picture that young star clusters are more likely to expand \citep{dellacroce_etal2024_expansion}.
Also, a clear trend of $\vRsigmaR$ with the distance from the center is observed within individual clusters suggesting that expansion dominates cluster dynamics in the outskirts.

The \w region was targeted by many studies that characterized the YSO population and spatial distribution. We complemented these studies by investigating the YSO kinematics within the region.
YSOs were found to trace young star clusters' expansion and, most probably, the parent gas bulk motion.
Finally, we characterized the candidate ionizing sources in H~II regions finding that at least one for each H~II region was assigned as a high-probable ($\geq 65\%$) cluster member.

Lastly, we further zoomed out to study the Perseus complex kinematics using its star clusters.
Six-dimensional phase space data were obtained from the latest \textit{Gaia} compilation coupled with large spectroscopic surveys (mainly for the LOS velocity component).
Star clusters trace the expansion rate reported by \citet{romanzuniga_etal2019} when looking at on-sky coordinates.
However, we found that such expansion is likely a projection effect due to different orbital velocities at slightly different Galactocentric distances.
Integrating the orbits of the three major structures in the complex (i.e., \lisca{I}, \lisca{II}, and W345), we traced their relative distance with time, concluding that: 
{\it i}) \lisca{I} is drifting away from W345, reaching a distance of
about 1.1 kpc in 100 Myr, before approaching it again;
{\it ii}) at the same time, \lisca{I} and \lisca{II} are getting closer, reaching their minimum relative distance in about $30-40$~Myr;
{\it iii}) \lisca{II} and \w are approaching each other:
in about 60 Myr they reach a distance of a few hundred parsecs, before slowly departing;
{\it iv}) we do not observe the Hubble-like expansion of the region suggested by \citet{romanzuniga_etal2019}. According to their reported rate, in 150 Myr the region should reach a size of about 5 kpc,
inconsistent with their orbit in the Galaxy.
In addition, backward orbit integration provides us with insights into the formation conditions of the Perseus complex: \lisca{I} and the \w region were at their minimum distance about 25 Myr ago of just a few hundred parsecs, while \lisca{II} formed further away, between $0.6-1$~kpc.

We also tested the role of spiral-arm perturbations in the orbit evolution since the Perseus complex spatially coincides with the Perseus spiral arm \citep[see e.g.,][]{reid_etal2019}.
The spiral arm perturbs the cluster orbits by dragging them toward higher-density regions, 
thus possibly keeping clusters closer for longer times when compared to the axisymmetric
case. We found this result to be fairly robust on short time scales ($\lesssim 100~$Myr) to varying density perturbation strength.

In summary, we presented a detailed characterization of the Perseus complex, starting from the clusters in the \w region up to its kinematics on large scales by progressively zooming out.
Particular attention was paid to complementing the numerous previous literature studies with kinematic data from \textit{Gaia} DR3. We showed indeed that kinematics (supplemented by photometric and spectroscopic data) is key to understanding the formation and evolution of large stellar complexes from cluster scales to Galactic ones.

\begin{acknowledgements}
We thank the anonymous referee for their valuable comments that helped improve the paper.
A.D.C. thanks L. Briganti, R. Pascale, L. Rosignoli, A. Mazzi, M. De Leo, and G. Ettorre for useful discussions.
A.D.C. and E.D. would like to thank A. Sills and S. Kamann for their valuable feedback and comments on the Ph.D. thesis, of which this paper is a part.
This work uses data from the European Space Agency (ESA) space mission \textit{Gaia}. \textit{Gaia} data are being processed by the \textit{Gaia} Data Processing and Analysis Consortium (DPAC). Funding for the DPAC is provided by national institutions, in particular, the institutions participating in the \textit{Gaia} Multi-Lateral Agreement (MLA).
This research or product makes use of public auxiliary data provided by ESA/Gaia/DPAC/CU5 and prepared by Carine Babusiaux.
RGB images in this work were processed through the astronomical image processing tool SIRIL (\url{https://siril.org/}).
Data underlying this article will be shared upon reasonable request to the corresponding author.
\end{acknowledgements}

\bibliographystyle{aa}
\bibliography{bibliographyfile}

\onecolumn
\appendix

\section{Perseus star cluster orbits in an axisymmetric potential \label{appendix:cluster_orbits}}

\begin{figure*}[!th]
    \centering
    \includegraphics[width=0.33\textwidth]{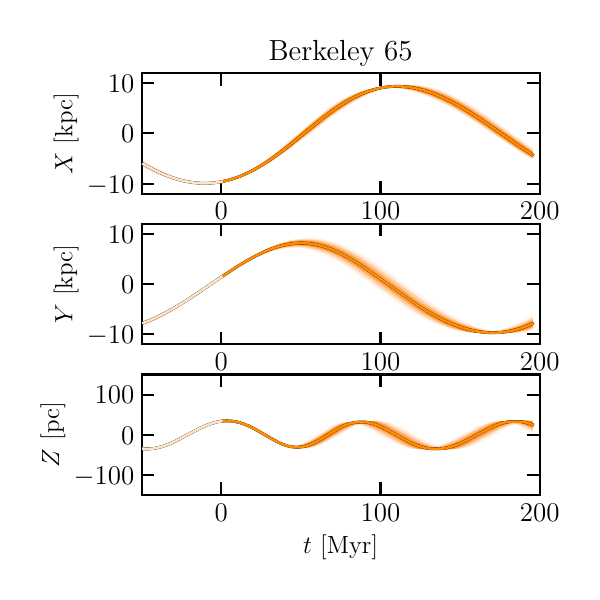}
    \includegraphics[width=0.33\textwidth]{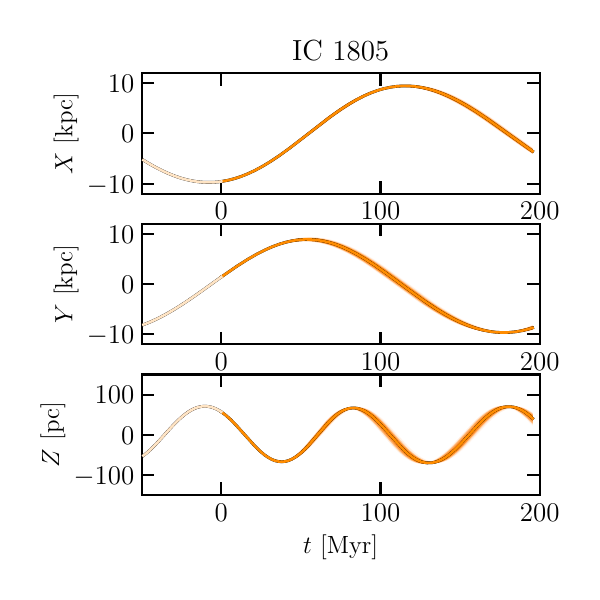}
    \includegraphics[width=0.33\textwidth]{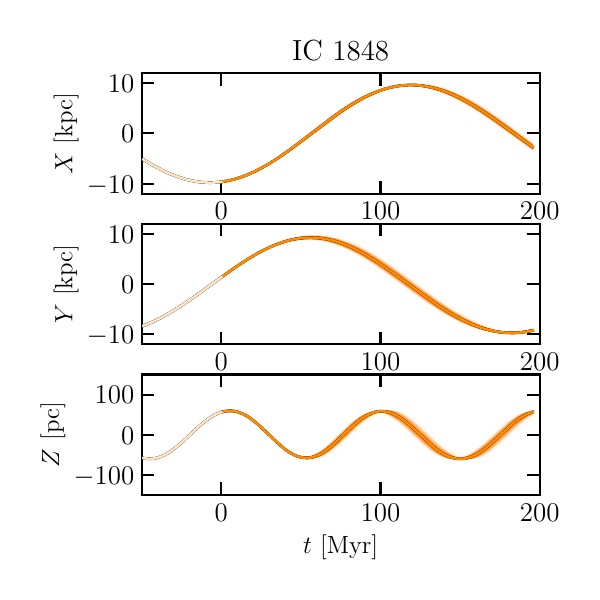}
    \includegraphics[width=0.33\textwidth]{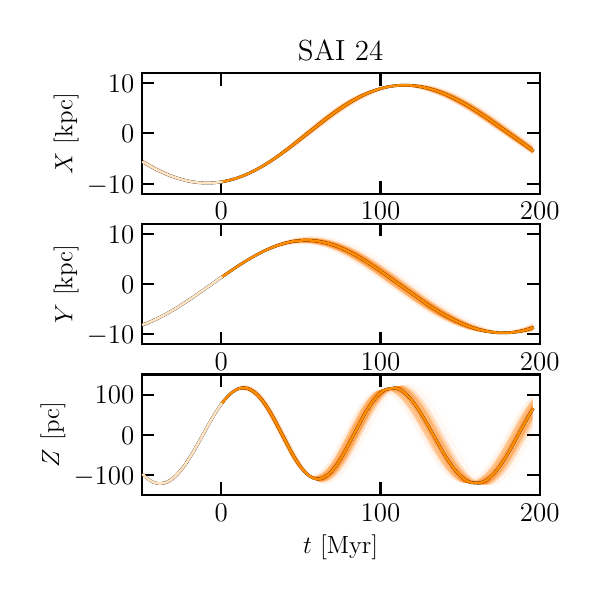}
    \includegraphics[width=0.33\textwidth]{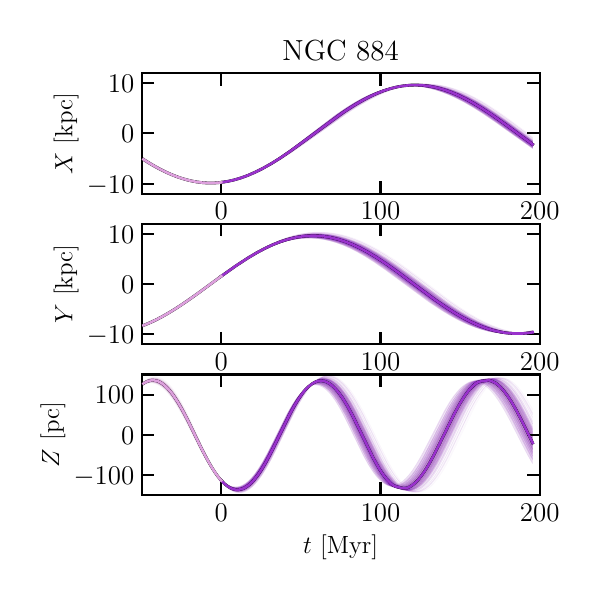}
    \includegraphics[width=0.33\textwidth]{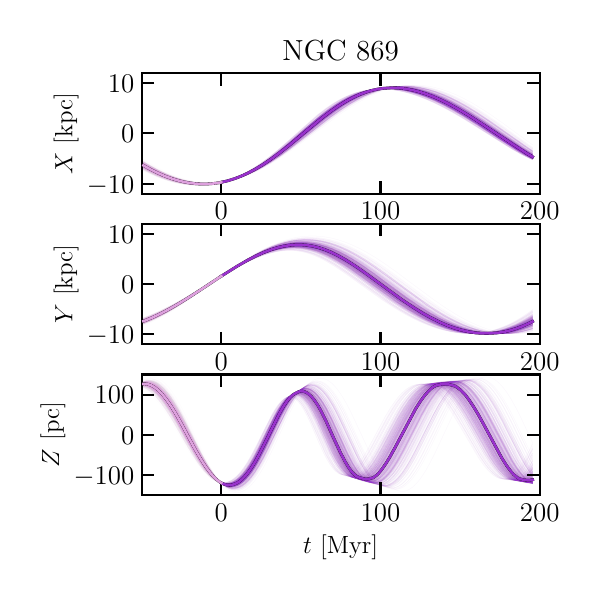}
    \includegraphics[width=0.33\textwidth]{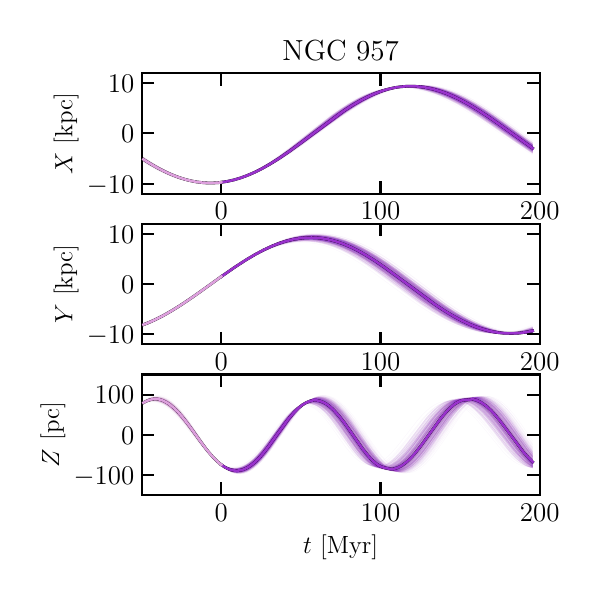}
    \includegraphics[width=0.33\textwidth]{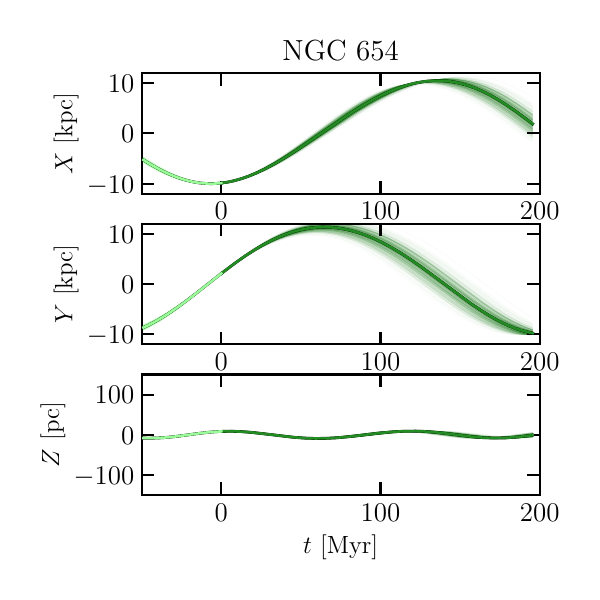}
    \includegraphics[width=0.33\textwidth]{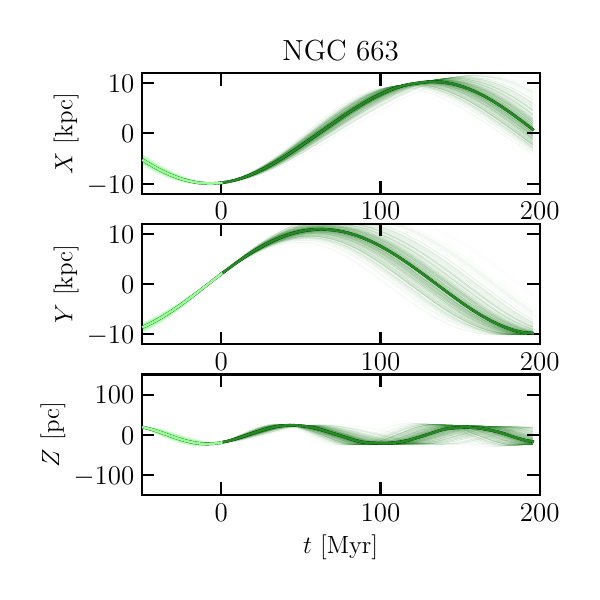}
    \caption{Temporal evolution of XYZ Galactocentric coordinates for the 15 star clusters in the Perseus complex. Temporal evolution for 500 orbit integrations is shown with background thin lines. Initial conditions were extracted according to the distance and $\vlos$ error distributions. 
    Thicker foreground lines depict the median over multiple integrations.
    Cluster orbits are color-coded according to the larger cluster agglomerate to which they belong: orange for \w complex, purple for \lisca{I}, and green for \lisca{II}.}
    \label{fig:orbits_starCluster_axisymemtricPotential}
\end{figure*}
\begin{figure*}[!th]
    \centering
    \includegraphics[width=0.33\textwidth]{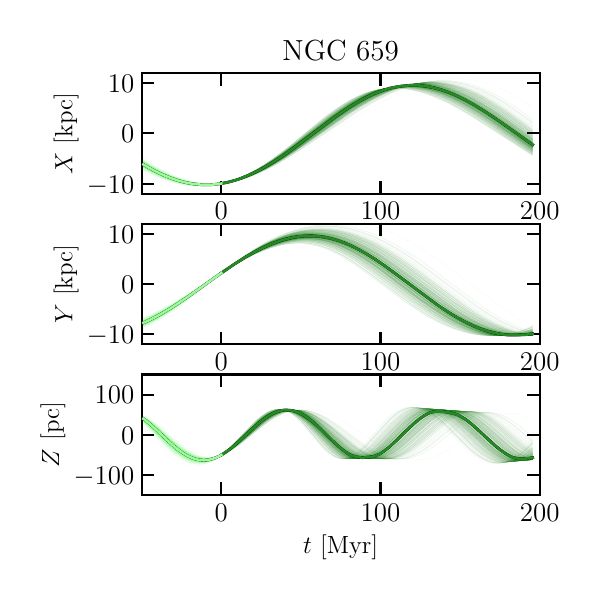}
    \includegraphics[width=0.33\textwidth]{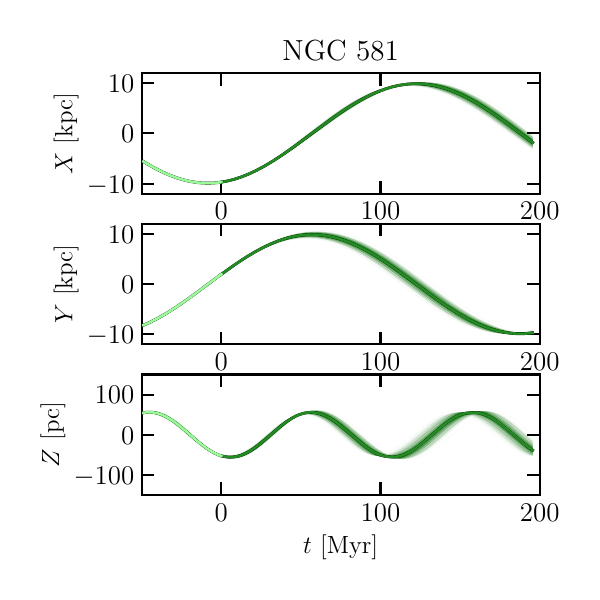}
    \includegraphics[width=0.33\textwidth]{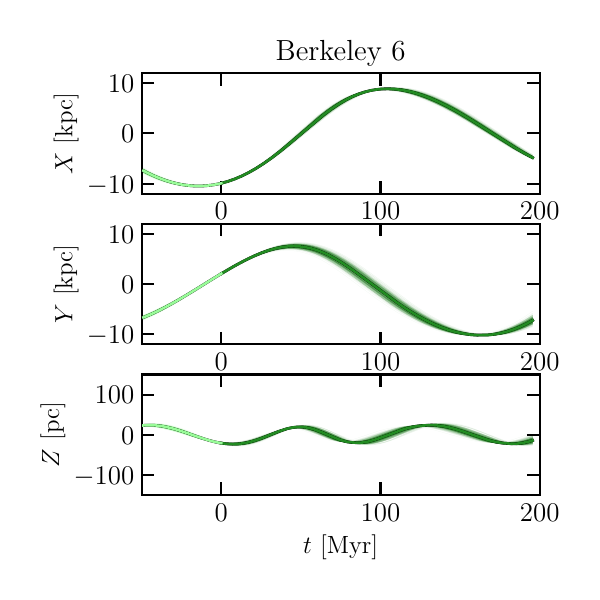}
    \includegraphics[width=0.33\textwidth]{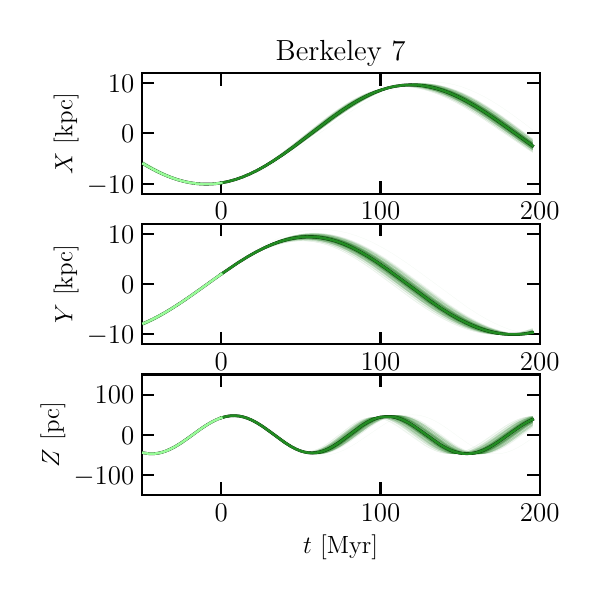}
    \includegraphics[width=0.33\textwidth]{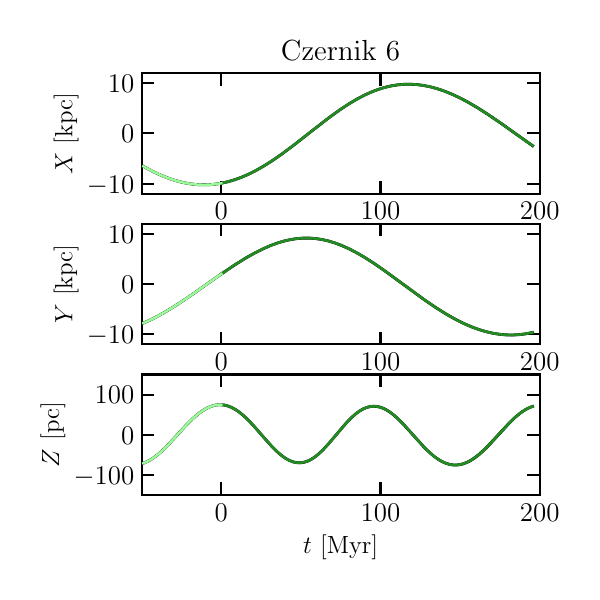}
    \includegraphics[width=0.33\textwidth]{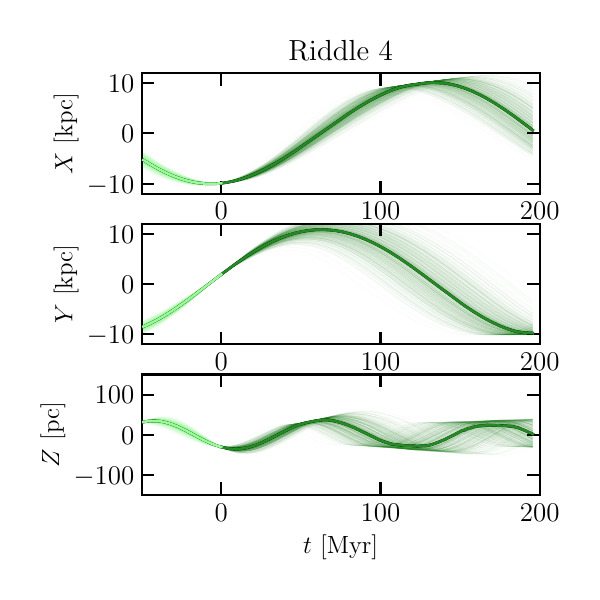}
    \caption{Figure~\ref{fig:orbits_starCluster_axisymemtricPotential} continues.}
    \label{fig:orbits_starCluster_axisymemtricPotential_continued}
\end{figure*}
This section presents the individual cluster orbits integrated within the axisymmetric \citet{mcmillan_etal2017} potential. The evolution of each cluster was followed for 200 Myr forward in time and 50 Myr backward.
Figures~\ref{fig:orbits_starCluster_axisymemtricPotential}, and \ref{fig:orbits_starCluster_axisymemtricPotential_continued} show the $XYZ$ Galacto-centric coordinates as a function of time for all 15 clusters. 
\FloatBarrier

\section{Testing the impact of the Galactic bar}
\label{appendix:bar_potential}
\begin{figure}[!th]
    \centering
    \includegraphics[width=0.5\textwidth]{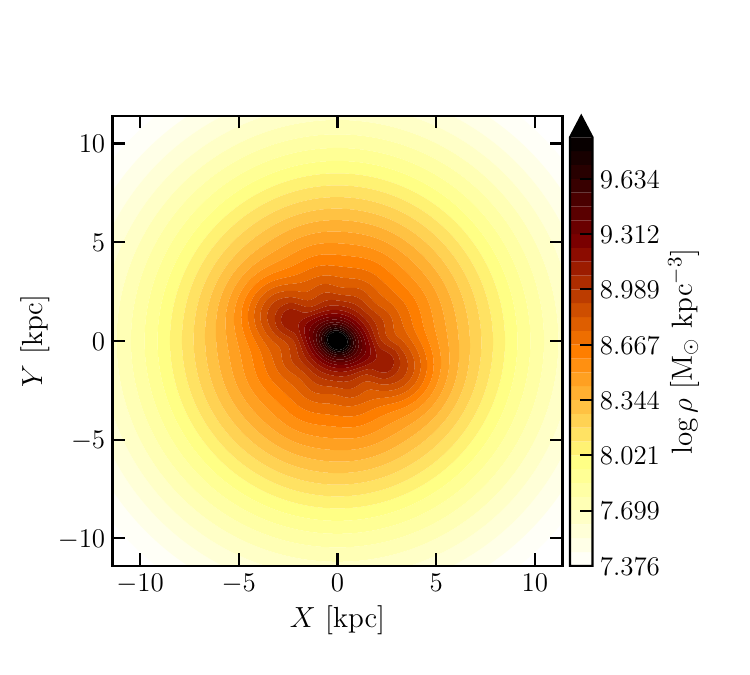}
    \caption{Density map on the XY Galactic plane (i.e., computed at $Z=0$) for a Galactic potential which includes a titled, rotating bar.}
    \label{fig:densityplot_barPotential}
\end{figure}
\begin{figure}[!th]
    \centering
    \includegraphics[width=0.5\textwidth]{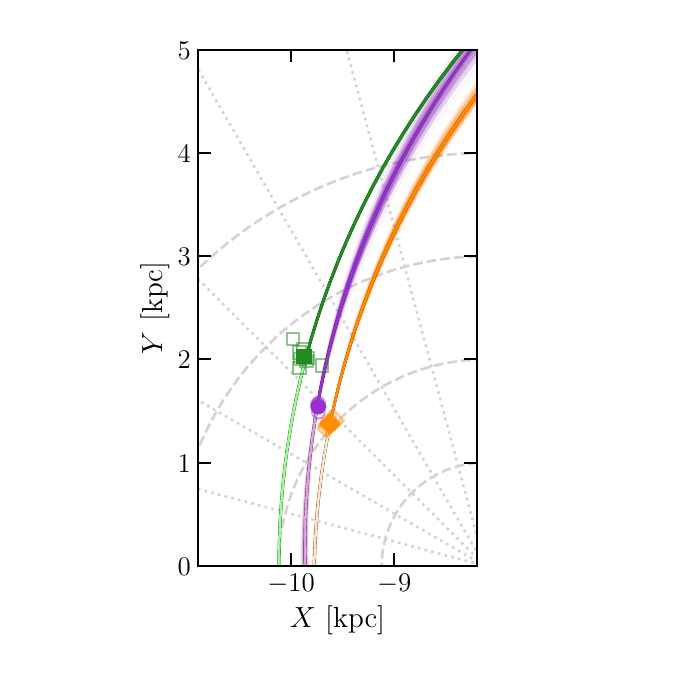}
    \caption{Orbits for the three stellar aggregates studied in this work on the Galactic plane. Orbits were computed within the potential model including the Galactic bar.}
    \label{fig:XYorbit_barPotential}
\end{figure}
\begin{figure}[!th]
    \centering
    \includegraphics[width=0.5\textwidth]{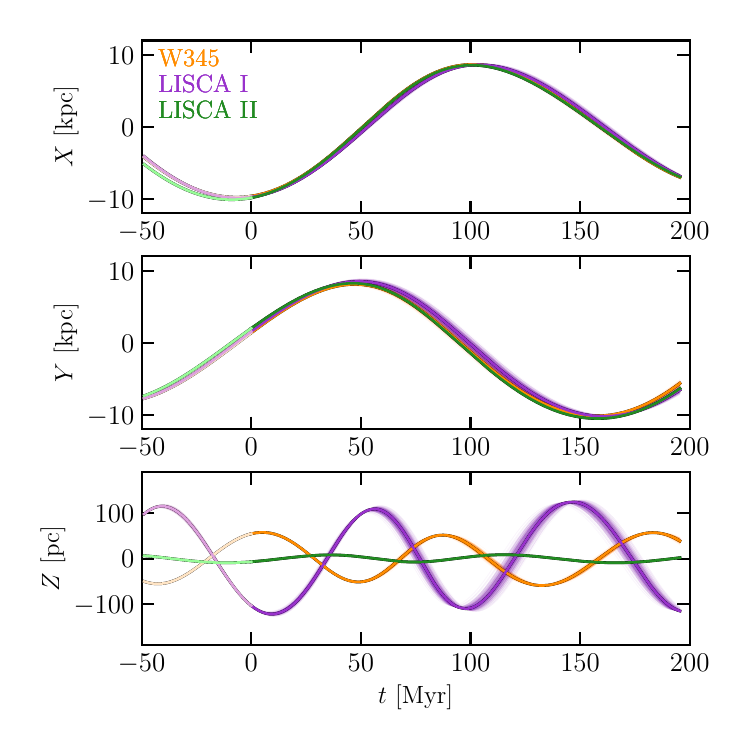}
    \caption{Galactocentric coordinates time evolution for the three stellar aggregates studied in this work. Orbits were computed within the potential model including the Galactic bar. Thin lines show different integrations from error distribution sampling of the initial conditions, while thicker ones are the median position at any given time.}
    \label{fig:XYZorbit_barPotential}
\end{figure}
\begin{figure}[!th]
    \centering
    \includegraphics[width=0.5\textwidth]{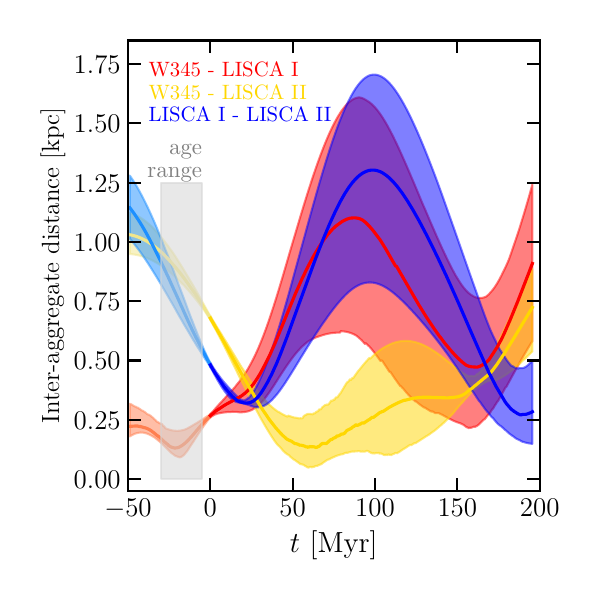}
    \caption{3D inter-cluster aggregate distance as a function of time. Orbits were computed in the Galactic potential which includes the bar.}
    \label{fig:interAggregateDistanceEvolution_barPotential}
\end{figure}

In this Section, we present the impact of the Galactic bar on the cluster aggregate orbits. In particular, we constructed the bar potential as defined in \citet{sormani_etal2022}. Briefly, the authors provided analytic formulae to match the $N$-body models by \citet{portail_etal2017}.
Such numerical models were in turn constrained to match the red clump stars density from a combination of infrared surveys, and the stellar kinematics in the bulge and the bar regions \citep{portail_etal2017}.
The analytical model includes three components: an X-shaped, a long, and a short bar \citep{sormani_etal2022}. The model was implemented within the \texttt{AGAMA} library\footnote{See the \texttt{Python} example scripts in the documentation: \url{https://github.com/GalacticDynamics-Oxford/Agama/tree/master/py}.} and considered as a perturbation of the \citet{mcmillan_etal2017} potential (as done for the spiral arms, see Section~\ref{sec:orbits:spiral_potential}). Figure~\ref{fig:densityplot_barPotential} shows the 3D density computed on the Galactic plane ($Z=0$) constructed for such MW potential. The bar is roughly confined in the central 5~kpc, although resonances can strongly perturb orbits well outside that region. We assumed a tilt angle of $-25^\circ$ to the positive direction of the $X$ axis.

To quantitatively assess the impact of the Galactic bar, we then performed the same analysis as Section~\ref{sec:orbits_axysimmetric_potential} but for the MW potential model with the bar. For the purposes of orbit integration, we assumed $\Omega_{\rm p,\,bar} = 37.5~$km~s$^{-1}$~kpc$^{-1}$ \citep{sormani_etal2022}.
Figures~\ref{fig:XYorbit_barPotential} and \ref{fig:XYZorbit_barPotential} present the orbits in the MW model with the bar. In particular, Fig.~\ref{fig:XYorbit_barPotential} shows the orbits projected onto the Galactic plane, while in Fig.~\ref{fig:XYZorbit_barPotential} the time evolution (both forward and backward) of the Galactocentric coordinates are shown.
Finally, the impact of the bar on the 3D relative distances of the aggregates is presented in Fig.~\ref{fig:interAggregateDistanceEvolution_barPotential}. As expected, the results are remarkably similar to those obtained in the purely axisymmetric case (see Fig.~\ref{fig:interAggregateDistances_axisymmetricPot}), thus confirming that the Galactic bar has a negligible effect on the Perseus cluster orbits compared to the local spiral arm structure.

\FloatBarrier
\section{Exploring different $f$ values \label{appendix:different_spiral_models}}
\begin{figure*}[!th]
    \centering
    \includegraphics[width=0.25\textwidth]{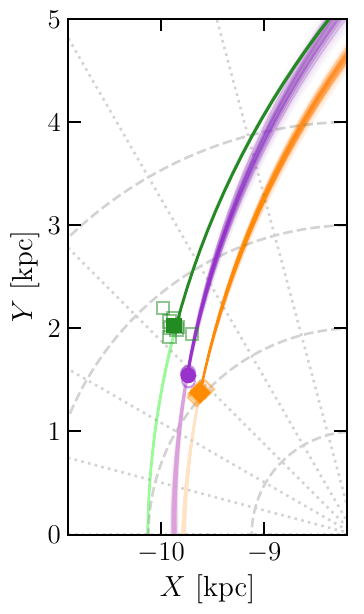}
    \includegraphics[width=0.25\textwidth]{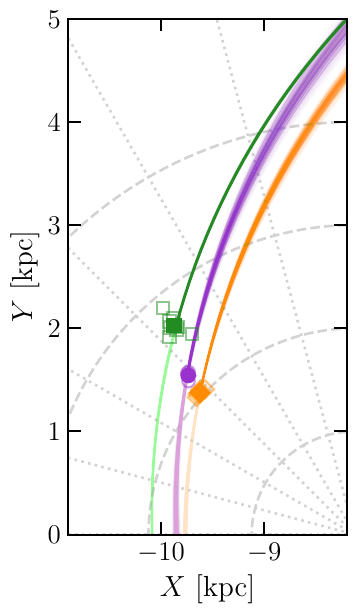}
    \includegraphics[width=0.25\textwidth]{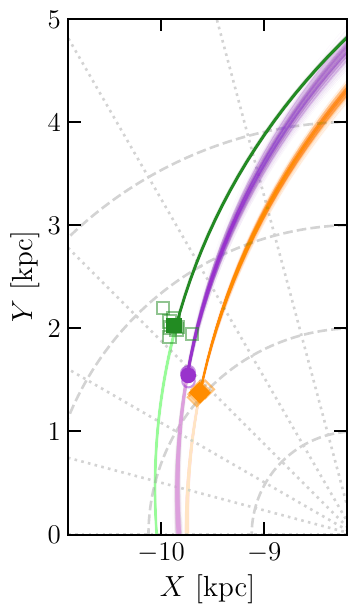}
    \caption{Stellar aggregate orbits on the Galactic plane for three values of $f$. From left to right $f = 0.1,\,0.3,\,0.5$, resulting in stronger density and potential perturbations. Different curves are orbits for different initial conditions extracted according to observed error distributions. \w, \lisca{I}, and \lisca{II} orbits are in orange, purple, and green respectively. Present-day positions are also shown with filled symbols.}
    \label{fig:XYproj_spiralModels_AppendixB}
\end{figure*}
\begin{figure*}[!th]
    \centering
    \includegraphics[width=0.32\textwidth]{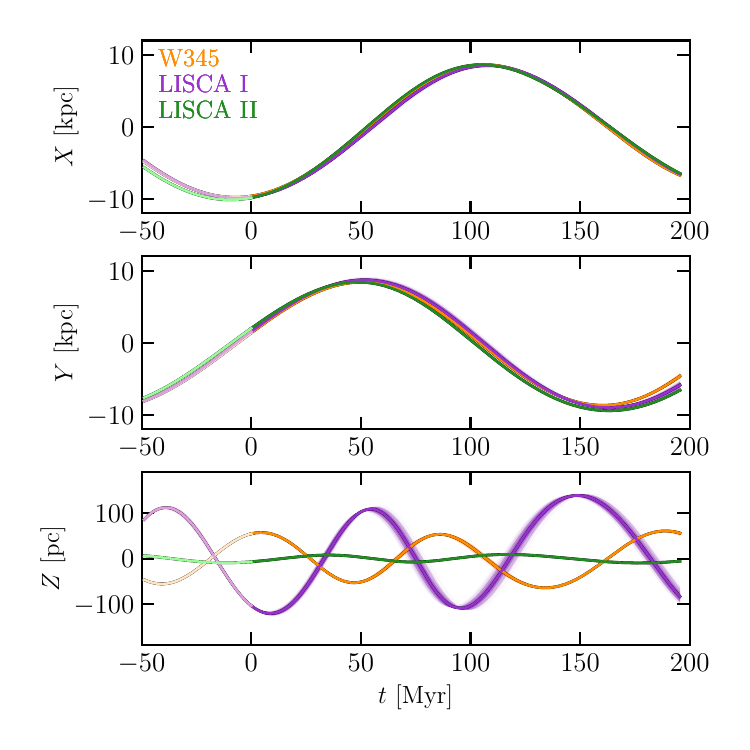}
    \includegraphics[width=0.32\textwidth]{plots/newCoords/XYZproj_clustersAggregate_f03.pdf}
    \includegraphics[width=0.32\textwidth]{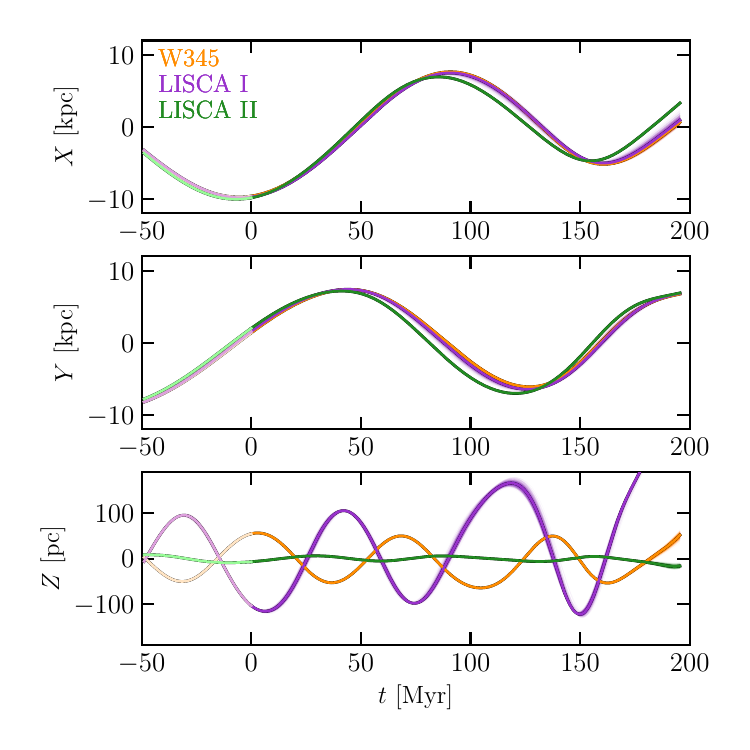}
    \caption{Temporal evolution of Galactocentric coordinates for the three cluster aggregates integrated in different spirally-perturbed potential: from left to right $f = 0.1,\,0.3,\,0.5$.
    Darker colors show the forward integration for about 200~Myr, whereas lighter ones show the backward integration. Different colors depict different aggregates: the \w complex in orange, \lisca{I} in purple, \lisca{II} in green.}
    \label{fig:XYZevolution_spiralModels_AppendixB}
\end{figure*}
\begin{figure*}[!th]
    \centering
    \includegraphics[width=0.32\textwidth]{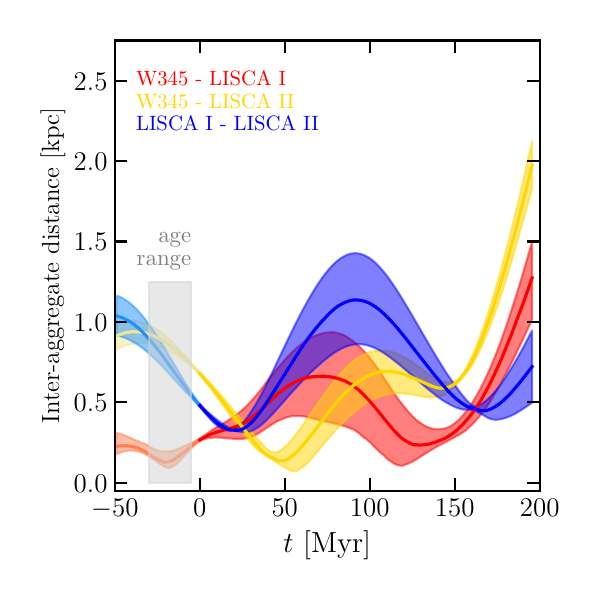}
    \includegraphics[width=0.32\textwidth]{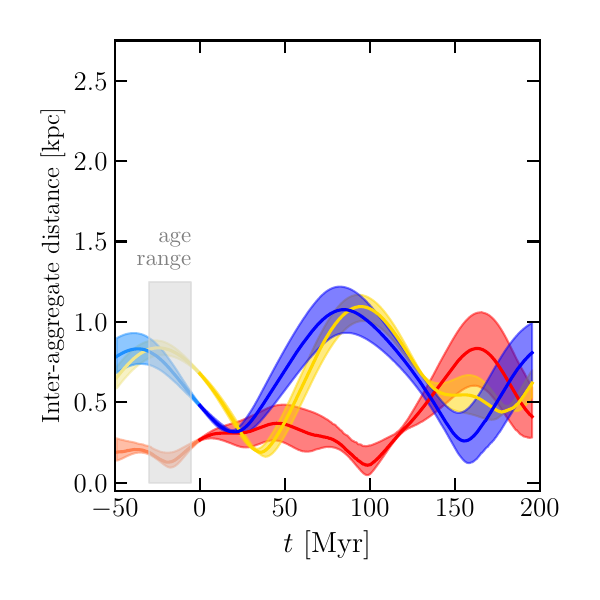}
    \includegraphics[width=0.32\textwidth]{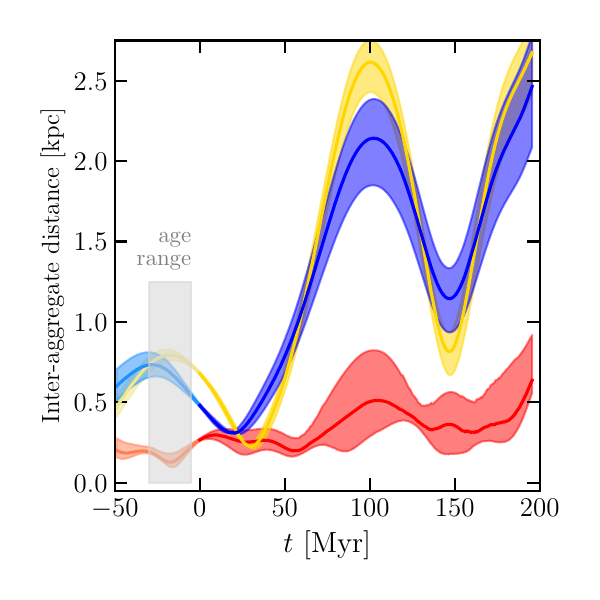}
    \caption{Inter-aggregate distances evolution for different spirally-perturbed potentials. From left to right $f = 0.1,\,0.3,\,0.5$.
    Mind the different scales on the $y$-axis. The gray shaded areas mark the cluster ages range, from $5-30~$Myr.}
    \label{fig:distanceEvolution_spiralModels_AppendixB}
\end{figure*}
In this section, we show the cluster aggregates orbits for different non-axisymmetric potentials. Among the different integrations only $f$ was changed exploring $f=0.1, 0.3, 0.5$.
Figure~\ref{fig:XYZevolution_spiralModels_AppendixB} presents the $XY$ projections of the aggregate orbits on the Galactic plane ($f$ is increasing rightward).
Similarly, Fig.~\ref{fig:XYZevolution_spiralModels_AppendixB} presents the evolution with time of the $XYZ$ coordinates for the three cluster aggregates.
As could be seen, the stronger the density perturbation (i.e., increasing $f$) the more bend the orbits, which tend to follow the spiral structure.
Finally, the implications of stronger spiral perturbations on the inter-aggregate distances are shown in Fig.~\ref{fig:distanceEvolution_spiralModels_AppendixB}.

\end{document}